\documentclass[twocolumn,showpacs,showkeys]{revtex4}

\usepackage{amsfonts}
\usepackage{amsmath}
\usepackage{graphicx}

\begin{document}


\title{Motion in gauge theories of gravity}


\author{Romualdo Tresguerres}
\email[]{romualdotresguerres@yahoo.es}
\affiliation{Instituto de F\'isica Fundamental\\
Consejo Superior de Investigaciones Cient\'ificas\\ Serrano 113 bis, 28006 Madrid, SPAIN}

\date{\today}

\begin{abstract}
A description of motion is proposed, adapted to the composite bundle interpretation of Poincar\'e Gauge Theory. Reference frames, relative positions and time evolution are characterized in gauge-theoretical terms. The approach is illustrated by an appropriate formulation of the familiar example of orbital motion induced by Schwarzschild spacetime.
\end{abstract}

\maketitle

\section{Introduction}

Any conception of motion is based on an underlying geometrical framework describing the structure assumed for spacetime. Typically, Newton's idea of absolute motion rests on absolute Euclidean space and absolute time, whose existence is postulated even though, in view of Galilean symmetry, only relative motion referred to inertial frames is observable. Moreover, Einstein's relative motion as depicted by Special Relativity is supported on Minkowski spacetime, and the geodesic motion of free particles and of light rays in General Relativity takes place on the pseudo-Riemannian metric manifold $(M\,,g)$ with Lorentzian signature where the theory is formulated.

Nevertheless, to the author's knowledge, no description of motion exists which is properly adapted to the characteristic geometrical structure of gauge theories of gravitation. Certainly, a reason for it can be found in the fact that the considerable effort made along the years in the construction of such gravitational theories of the Yang-Mills type \cite{Utiyama:1956sy}--\cite{Obukhov:2006ge} has produced a number of quite different proposals. Being all of them gauge theoretical approaches, they have in common the fiber bundle nature \cite{Daniel:1979ez}--\cite{Goeckeler}, but no further agreement exists, neither in the choice of the structure group nor in the particular bundle structure invoked by each one.

Einstein's General Relativity can be interpreted, at least in principle, in terms of tensor quantities defined on a tensor bundle associated to the principal bundle of (tangent) linear  frames LM with $GL(4\,,\mathbb{R})$ as structure group \cite{Mielke:1987gk}--\cite{Socolovsky:2011mv}. Gauge theories of gravitation such as Metric Affine Gravity (MAG) or Poincar\'e Gauge Theory (PGT) proposed by Hehl et al. \cite{Hehl:1974cn}--\cite{Obukhov:2006ge} are formulated on bundles of tensor-valued differential forms associated respectively to the principal bundle of affine frames $A(M)=LM\times\mathbb{R}^4$ or to that of Poincar\'e frames resulting from replacing the bundle $LM$ of linear frames with the bundle $FM$ of $SO(3\,,1)$ orthonormal frames \cite{Hehl:1995ue} \cite{Mielke:1987gk} \cite{Bleecker:1981me}. Lord \cite{Lord:1986xi}--\cite{Lord:1988nd} proposes the principal bundle structure $G(G/H\,,H)$ with $G=$Poincar\'e and $H=$ the Lorentz subgroup $SO(3\,,1)$, where he accommodates his particular conception of gauge transformations as non-vertical automorphisms. Sardanashvily \cite{Sardanashvily} --\cite{Sardanashvily:2005mf} introduces composite fiber bundles $LM\rightarrow LM/H\rightarrow M$, with whose help he explains the spontaneous symmetry breaking from the structure group $GL(4\,,\mathbb{R})$ of the linear frame bundle $LM$ to $H=$the Lorentz group, allowing to interpret the tetrad variables as Goldstone fields \cite{Ivanenko:1983vf} \cite{Sardanashvily:2005mf}--\cite{Trautman:1979cq}. Finally, the author of the present paper proposed in Ref.\cite{Tresguerres:2002uh} a different composite fiber bundle structure $P\rightarrow P/H\rightarrow M$ built from the principal bundle $P(M\,,G)$ with $G=$Poincar\'e and $H=$Lorentz, suitable to deal with gauge theories of gravitation such as PGT or MAG, where gravity is considered as the gauge theory of a group including translations \cite{Hehl:1974cn}--\cite{Obukhov:2006ge} \cite{Ali:2007hu}. In this scheme, translational symmetry is treated like ordinary gauge symmetries, and tetrads become identified with (nonlinear) translational connections with the right gauge transformations. Actually, it was proven in \cite{Tresguerres:2002uh} that the composite bundle structure provides a geometrical interpretation of nonlinear realizations (NLR's) \cite{Coleman:1969sm}--\cite{Tiemblo:2005sx}. See Appendix A.

The aim of the present work is to show that, at least in the particular case of composite fiber bundles proposed by the author as the suitable geometrical framework for Poincar\'e Gauge Theory, a description of motion exists which is modeled by its necessary adaptation to the bundle structure underlying this gauge theory of gravitation.

The paper is organized as follows. In Sections II--IV, we review the main features of the Poincar\'e composite bundle structure constituting the necessary support for the subsequent development. We include several improvements and extensions as compared with the original exposition of Ref.\cite{Tresguerres:2002uh}. In particular, a 5-dimensional matrix representation of the Poincar\'e group is introduced. In Section V, we construct the basis vectors of the $P$-associated 5-D vector bundle, assuming the vector fifth component to play the role of an origin. In Sections VI and VII, we discuss the bundle tangent space, where we pay special attention to a certain set of four tangent vectors. After enlarging them with a point, we propose to regard the resulting 5-dimensional set as a Poincar\'e frame, consisting of the basis components of a 5-D vector representation space of the Poincar\'e group. Then, the composite bundle structure allows to define in a natural way an invariant point-like quantity ${\buildrel {_\Sigma}\over p}$ which --in a certain sector of the tangent space-- provides a local description of positions referred to the local Poincar\'e frames previously built.

Section VIII is devoted to deal with different aspects of time evolution as induced by a (timelike) vector field ${\buildrel {_\Sigma}\over u}$ constituting the generator of the flow of events. So, we perform a simultaneity defining foliation (related to a certain time connection) and we examine the various appearances of the timelike vector field --playing the role of both, time and fourvelocity-- when expressed in terms of fields characteristic for the underlying bundle structure. In particular, we find that the fourvelocity components take the form of a Poincar\'e covariant generalization of the Lie derivative of certain translational fields resembling the spacetime coordinates of Special Relativity. Finally, after reformulating geodesic motion as induced by ${\buildrel {_\Sigma}\over u}$ on ${\buildrel {_\Sigma}\over p}$, in Section IX we apply our view to test particles orbiting on a Schwarzschild spacetime. The Conclusions are found in Section X.

\section{Composite fiber bundles}

\subsection{Rearranging a principal bundle manifold}

Ordinary gauge theories of internal groups $G$ have the geometrical structure of principal bundles $P(M\,,G)$, where the base space $M$ represents spacetime and the connections --defining horizontality on the bundle-- manifest themselves physically as gauge potentials describing interactions; matter fields are defined on associated bundles \cite{Daniel:1979ez}--\cite{Goeckeler} \cite{Trautman:1970cy}. However, this bundle scheme fails in accomodating gauge theories of spacetime groups. In Ref.\cite{Tresguerres:2002uh}, the author invoked a modified version of the composite fiber bundles introduced by Sardanashvily \cite{Sardanashvily}--\cite{Sardanashvily:2005mf} as the suitable topological structure underlying gauge theories of gravity such as PGT or MAG \cite{Hehl:1974cn}--\cite{Obukhov:2006ge}. Let us recall, with slight modifications, the bundle features established in \cite{Tresguerres:2002uh} which are relevant for the present paper. For what follows, see also \cite{Tiemblo:2005sx}, as much as \cite{Kobayashi:1963}, pags. 54 and 57.

In accordance with Sardanashvily's conception of composite bundles \cite{Sardanashvily}--\cite{Sardanashvily:2005mf}, let us enunciate the following theorem, mainly based on propositions 5.5 and 5.6 of the book \cite{Kobayashi:1963} by Kobayashi and Nomizu:

Let $\pi _{_{PM}}:P\rightarrow M$ be a principal fiber bundle whose structure group $G$ is reducible to a closed subgroup $H\subset G$. Then, the principal bundle can be rearranged as a composite manifold
\begin{equation}
\pi _{_{\Sigma M}}\circ\pi _{_{P\Sigma}}:P\rightarrow\Sigma\rightarrow M\label{compbundle01}
\end{equation}
with an {\it intermediate space} $\Sigma =P/H$, in such a way that
\begin{equation}
\pi _{_{P\Sigma}}:P\rightarrow\Sigma\label{compbundle02}
\end{equation}
is a principal subbundle of $P$ with structure group $H$, and
\begin{equation}
\pi _{_{\Sigma M}}:\Sigma\rightarrow M\label{compbundle03}
\end{equation}
is a $P$-associated bundle with typical fiber $G/H$ and structure group $G$, being $\pi _{_{\Sigma M}}\circ\pi _{_{P\Sigma}} =\pi _{_{PM}}$ as established in (\ref{compbundle05}) below. Global sections $s_{_{M\Sigma}}: M\rightarrow\Sigma $ of (\ref{compbundle03}) exist, playing the role of Goldstone-like fields.

The proof obeys the following scheme \cite{Kobayashi:1963}. Given a principal fiber bundle $P(M\,,G)$ whose structure group $G$ possesses a closed subgroup $H$ of $G$, one can build the $P$-associated bundle $\Sigma (M\,,G/H\,,G\,,P)$ (see pg. 54 of \cite{Kobayashi:1963}), having the same structure group $G$, standard fibre $G/H$ and base space $M$. Proposition 5.5. of \cite{Kobayashi:1963} proves that the total space of this associated bundle $\Sigma $ can be identified with the quotient space $P/H$ of $P$ by the right action of $H$ on $P$, thus implying that $P(\Sigma\,,H)$ is a principal fibre bundle over the {\it (intermediate) base space} $\Sigma = P/H$, with structure group $H$ and with well defined projection $\pi _{_{P\Sigma}}:P\rightarrow\Sigma\,$. On the other hand, according to Proposition 5.6 of \cite{Kobayashi:1963}, the necessary and sufficient condition for the structure group $G$ of $P(M\,,G)$ to be reducible to the closed subgroup $H\subset G$ is that the associated bundle $\Sigma = P/H$ admits a cross section $s_{_{M\Sigma}}: M\rightarrow\Sigma $. Provided such sections exist, there is a one to one correspondence between them and the reduced subbundles of $\pi _{_{P\Sigma}}:P\rightarrow\Sigma$ consisting (see the proof of 5.6 in \cite{Kobayashi:1963}) of the set of points $u\in P$ such that
\begin{equation}
\pi _{_{P\Sigma}}(u) =s_{_{M\Sigma}}\circ\pi _{_{PM}}(u)\,.\label{compbundle04}
\end{equation}
Taking into account the section condition $\pi _{_{\Sigma M}}\circ s_{_{M\Sigma}}= id_{_M}$, where $id_{_M}$ is the identity transformation of $M$, multiplication of both members of (\ref{compbundle04}) by $\pi _{_{\Sigma M}}$ yields the decomposition
\begin{equation}
\pi _{_{PM}}=\pi _{_{\Sigma M}}\circ\pi _{_{P\Sigma}}\label{compbundle05}
\end{equation}
of the total bundle projection $\pi _{_{PM}}$ into partial projections. Accordingly, the principal bundle $\pi _{_{PM}}:P\rightarrow M$ transforms into the composite bundle (\ref{compbundle01}), where we distinguish the bundle sectors (\ref{compbundle02}) and (\ref{compbundle03}), characterized by the partial projections $\pi _{_{P\Sigma}}$ and $\pi _{_{\Sigma M}}$ respectively. Sector (\ref{compbundle03}), that is $\Sigma =P/H$, is locally isomorphic to the Cartesian product $\Sigma\simeq M\times G/H$. Its elements trivialize locally as $(x\,,\xi )$, with $\xi$ coordinatizing the fiber branches $G/H$ attached to $x\in M$.

In view of the composite bundle structure (\ref{compbundle01})--(\ref{compbundle03}), the $G$-diffeomorphic fibers of $P(M\,,G)$ projecting to the bundle base space $M$ become {\it bent} into two sectors, corresponding to the fibers of $P(\Sigma\,,H)$ and $\Sigma\simeq M\times G/H$ respectively, in such a way that the $H$-diffeomorphic fiber branches $\pi _{_{P\Sigma}}^{-1}(x\,,\xi\,)$ of $P(\Sigma\,,H)$ are attached to points $(x\,,\xi\,)$ of the fibred space $\Sigma\simeq M\times G/H$, playing the role of {\it base space} of the bundle sector $P(\Sigma\,,H)$. See Fig.1, where we include the coordinates of the Poincar\'e composite bundle to be considered later.
\begin{figure}[h]
\begin{center}
\includegraphics[totalheight=63mm,width=89mm]{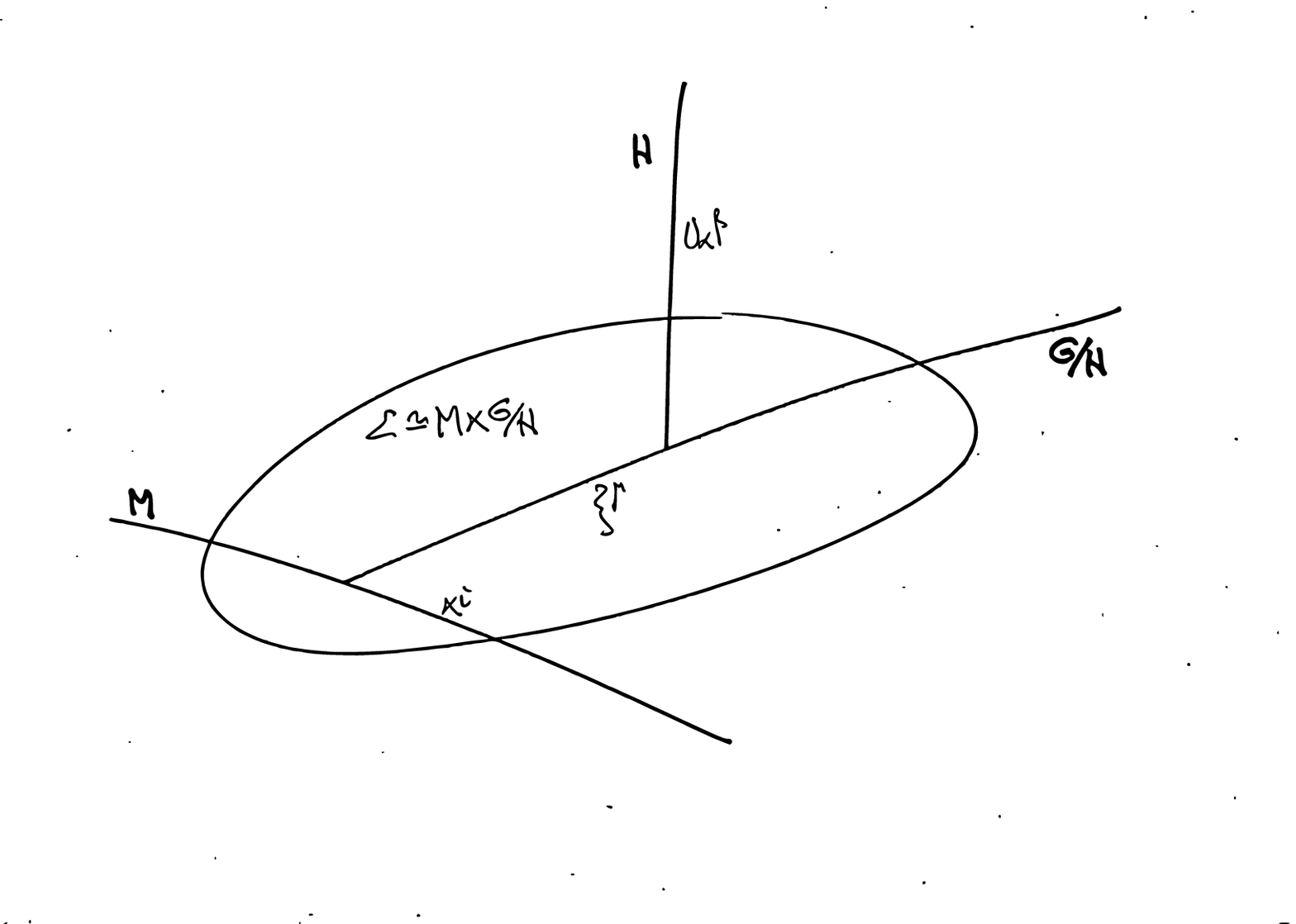}
\caption{\label{Fig1}Bundle manifold}
\end{center}
\end{figure}

In this scheme, the base space $M$ is not identified with spacetime. It rather plays the role of an auxiliary manifold underlying the true (Poincar\'e) spacetime variables, which are defined on the fibres (although they can be pulled back to $M$ and expressed as functions of coordinates $x\in M$).

\subsection{Sections and connections of composite bundles}

In correspondence with the decomposition (\ref{compbundle05}) of the bundle projection, we also decompose local sections $s_{_{MP}}:M\rightarrow P$ as
\begin{equation}
s _{_{MP}}=s _{_{\Sigma P}}\circ s_{_{M\Sigma}}\,,\label{sectdecomp}
\end{equation}
imposing that the respective section conditions, $\pi _{_{PM}}\circ s _{_{MP}}=\pi _{_{\Sigma M}}\circ s_{_{M\Sigma}}= id_{_M}$ and $\pi _{_{P\Sigma }}\circ s _{_{\Sigma P}}= id_{_\Sigma}$, are satisfied; see \cite{Tresguerres:2002uh}. Next we express the general sections $s_{_{MP}}$ in terms of the right action of elements $\tilde{g}$ of the structure group $G$ on suitable zero sections denoted as $\sigma _{_{MP}}$, that is
\begin{equation}
s_{_{MP}} =R_{\tilde{g}}\circ\sigma _{_{MP}}\,,\qquad \tilde{g}\in G\,.\label{locsect1}
\end{equation}
If one factorizes the arbitrary group elements $\tilde{g}\in G$ in (\ref{locsect1}) into group products of the form
\begin{equation}
\tilde{g} =b\cdot a\,,\qquad a\in H\,,\quad b\in G/H\,,\label{condit1}
\end{equation}
consistently with the decomposition of the bundle manifold $P(M\,,G)$ into the bundle sectors (\ref{compbundle02}) and (\ref{compbundle03}) (whose fibres are diffeomorphic to $H$ and $G/H$ respectively), then one can rewrite the sections in the r.h.s. of (\ref{sectdecomp}), in analogy to (\ref{locsect1}), as
\begin{eqnarray}
s_{_{\Sigma P}} &=& R_a\circ\sigma _{_{\Sigma P}}\,,\qquad a\in H\,,\label{locsect2}\\
s_{_{M\Sigma}} &=& R_b\circ\sigma _{_{M\Sigma}}\,,\qquad b\in G/H\,,\label{locsect3}
\end{eqnarray}
respectively. Assuming
\begin{equation}
R_b\circ\sigma _{_{MP}} =\sigma _{_{\Sigma P}}\circ R_b\circ\sigma _{_{M\Sigma}}\,,\label{condit2}
\end{equation}
eqs. (\ref{sectdecomp})--(\ref{condit2}) yield the (\ref{sectdecomp})-analogous relation
\begin{equation}
\sigma _{_{MP}}=\sigma _{_{\Sigma P}}\circ\sigma _{_{M\Sigma}}\,.\label{zerosectdecomp}
\end{equation}
We also use the factorization (\ref{condit1}) to build a connection form with two sectors, in such a way that it allows to describe separate horizontalities on both, the tangent spaces of $\Sigma\simeq M\times G/H$ and $P(\Sigma\,,H)$ respectively. We propose to rewrite the connection 1-form of ordinary principal bundles $P(M\,,G)$, namely
\begin{equation}
\omega =\tilde{g}^{-1}(\,d_{_P} +\pi _{_{PM}}^*\,A_{_M})\,\tilde{g}\,,\label{compconnform3}
\end{equation}
with the $G$-element $\tilde{g}=b\cdot a$ as given by (\ref{condit1}), as the connection 1-form of the total composite bundle $P\rightarrow\Sigma\rightarrow M$, that is
\begin{equation}
\omega =a^{-1}(\,d_{_P} +\pi _{_{P\Sigma}}^*\,\Gamma _{_{\Sigma}})\,a\,,\label{connform1}
\end{equation}
including the connection contribution
\begin{equation}
\Gamma _{_\Sigma} = b^{-1}(\,d_{_\Sigma} +\pi _{_{\Sigma M}}^*\,A_{_M})\,b\label{connform2}
\end{equation}
corresponding to the $\Sigma\rightarrow M$ sector. The objects (\ref{compconnform3}), (\ref{connform1}) and (\ref{connform2}) are built either with the differential $d_{_P}$ defined on the whole bundle manifold $P$ or with $d_{_\Sigma}$ defined on the {\it intermediate space} $\Sigma$, assuming that the different exterior derivatives of $b$ are related as
\begin{equation}
\pi _{_{P\Sigma}}^*\,d_{_\Sigma}b =d_{_P}b\,.\label{auxil01}
\end{equation}
(We use a notation taken from \cite{Nakahara}, pgs. 332 ff. In the following, we only indicate the manifolds where exterior derivatives are defined when necessary to avoid confusions.) Notice that (\ref{connform2}) coincides with the pullback of (\ref{connform1}) to the cotangent space $T^*(\Sigma )$ by means of a zero-section, namely
\begin{equation}
\Gamma _{_\Sigma} = \sigma _{_{\Sigma P}}^*\,\omega\,,\label{pbconn01}
\end{equation}
while the further pullback of (\ref{connform2}) itself to $T^*(M)$ yields
\begin{equation}
A_{_M} = \sigma _{_{M\Sigma}}^*\,\Gamma _{_\Sigma}\,,\label{pbconn02}
\end{equation}
so that, from (\ref{pbconn01}) and (\ref{pbconn02}), taking into account $\sigma _{_{M\Sigma}}^*\,\sigma _{_{\Sigma P}}^*\,\omega =\sigma _{_{MP}}^*\,\omega$ as deduced from (\ref{zerosectdecomp}) (see Appendix D), we find the standard expression
\begin{equation}
A_{_M} =\sigma _{_{MP}}^*\,\omega\label{pbconn03}
\end{equation}
of ordinary gauge potentials as the pullbacks of the connection 1-form (\ref{compconnform3}) to the base space $M$, in coincidence with gauge theories founded on ordinary principal bundles $P(M\,,G)$.

The role played by connection (\ref{connform1}) with (\ref{connform2}) in the characterization of distinguished horizontalities, defined respectively on (\ref{compbundle02}) and (\ref{compbundle03}), will be sudied in Sections IV and VI below for the particular case of $G=$Poincar\'e with $H=$Lorentz.

\subsection{Gauge transformations in composite bundles}

In \cite{Tresguerres:2002uh}, the author proposed an alternative to Lord's extension \cite{Lord:1986xi}--\cite{Lord:1988nd} of (active) gauge transformations to spacetime groups. In accordance with the standard definition \cite{Goeckeler} \cite{Atiyah:1978wi}, a gauge transformation on a principal fiber bundle $P(M\,,G\,)$ is a bundle automorphism $\lambda :P\rightarrow P$ satisfying two conditions, namely: it must commute with the right action of $G$ as
\begin{equation}
\lambda\circ R_g (u\,) =\,R_g\circ\lambda (u\,)\,,\label{gautrans01}
\end{equation}
(in such a way that it maps fibers to fibers), and in addition, it has to be a vertical automorphism, that is
\begin{equation}
\pi\circ\lambda (u\,) =\,\pi (u\,)\,,\label{gautrans02}
\end{equation}
so that no action is induced by it on the base space $M$, since both, $u$ and $\lambda (u\,)$, belong to the same fiber.

This conception of $\lambda$ derives in a natural way from gauge theories of internal groups, appropriate to describe any forces other than gravity. However, as observed by E. A. Lord \cite{Lord:1986xi}--\cite{Lord:1988nd}, the characterization of gauge transformations as vertical bundle automorphisms fails for gauge theories of spacetime groups: mainly for those based on groups including translations, such as PGT and MAG, where the group action on spacetime positions, usually represented by coordinates, must be taken into account. Accordingly, in order to extend to spacetime groups the validity of gauge transformations, Lord proposed to renounce to the verticality condition (\ref{gautrans02}), thus allowing the existence of nonvertical gauge transformations inducing diffeomorphisms on $M$. As a consequence of his view on gauge transformations, he suggested the principal bundle $G(G/H\,,H\,)$ as the general fiber bundle structure of gauge theories of gravitation. According to him, choosing for instance $G=$Poincar\'e and $H=$Lorentz, the base space $G/H$, identical with the parameter space of translations, plays the role of spacetime.

The author's own approach \cite{Tresguerres:2002uh} to gauge transformations, founded on composite fiber bundles, conciliates the standard definition with a suitable modification of Lord's point of view. Indeed, the most unsatisfactory aspect of Lord's bundle proposal is its lack of true translational connections, which cannot occur in the absence of an underlying base space $M$ to which standard translational fibers $G/H$ are projected. (This avoids to treat translations in the same way as the remaining symmetries. Actually, in theories like PGT and MAG \cite{Hehl:1974cn}--\cite{Obukhov:2006ge}, the --maybe modified, nonlinear-- translational connections are expected to play the role of tetrads.) The resort to composite bundles allows to have available translational connections and, at the same time, to restore the verticality condition (\ref{gautrans02}) over $M$, in a way compatible with induced spacetime transformations on the {\it intermediate space} $\Sigma$, as we will show immediately.
In \cite{Tresguerres:2002uh}, we justified the choice of $\lambda$ as the left action $L_g$ of group elements $g\in G$, local in the sense that $g=g(x)$, $x\in M$. Accordingly, identifying $u\in P$ with values of sections (\ref{locsect1}), gauge transformations $u' =\lambda (u)$ take the form
\begin{equation}
s_{_{MP}}'(x) = L_g\circ s_{_{MP}}(x)\,.\label{secttrans1}
\end{equation}
Making use of (\ref{locsect1}) and (\ref{condit1}), we find
\begin{equation}
s_{_{MP}}(x) =R_{\tilde{g}}\circ\sigma _{_{MP}}(x) = R_a\circ R_b\circ\sigma _{_{MP}}(x)\,.\label{smp1}
\end{equation}
Let us now define
\begin{equation}
\sigma _{\xi}(x) :=\sigma _{_{\Sigma P}}\circ s_{_{M\Sigma}}(x)\,,\label{sigmaxi}
\end{equation}
which constitutes a composite section $\sigma _{\xi}:M\rightarrow\Sigma\rightarrow P$ defined from the total and zero sections in (\ref{locsect2}) and (\ref{locsect3}) respectively. In view of (\ref{locsect3}) and (\ref{condit2}), from (\ref{sigmaxi}) follows
\begin{equation}
\sigma _{\xi}(x)=R_b\circ\sigma _{_{MP}}(x)\,,\label{sigmaxibis}
\end{equation}
the $\xi$ in $\sigma _{\xi}(x)$ standing for the parameters labelling the elements $b\in G/H$ displayed as $R_b$ in (\ref{sigmaxibis}). From (\ref{smp1}) and (\ref{sigmaxibis}) follows
\begin{equation}
s_{_{MP}}(x) = R_a\circ \sigma _{\xi}(x)\,.\label{smp2}
\end{equation}
In analogy to (\ref{smp1}) and (\ref{smp2}), we also have
\begin{eqnarray}
s_{_{MP}}'(x) &=& R_{\tilde{g}\,'}\circ\sigma _{_{MP}}(x)\nonumber\\
&=& R_{a\,'}\circ R_{b\,'}\circ\sigma _{_{MP}}(x)\nonumber\\
&=& R_{a '}\circ \sigma _{\xi '}(x)\,.\label{smp3}
\end{eqnarray}
Then, replacing (\ref{smp2}) and (\ref{smp3}) in (\ref{secttrans1}), we get $L_g\circ R_a\circ \sigma _{\xi}(x) = R_{a '}\circ \sigma _{\xi '}(x)$, implying
\begin{equation}
L_g\circ\sigma _\xi (x)=R_h\circ\sigma _{\xi '}(x)\,,\label{secttrans2}
\end{equation}
where
\begin{equation}
h:=a'\cdot a^{-1}\,,\label{hdef}
\end{equation}
as in (\ref{nlr03}). Notice that, acting with $\sigma _{_{\Sigma P}}$ on both members of (\ref{compbundle04}) and using (\ref{sigmaxi}), we get
\begin{eqnarray}
\sigma _{_{\Sigma P}}(x\,,\xi\,) &:=& \sigma _{_{\Sigma P}}\circ\pi _{_{P\Sigma}}(u)\nonumber\\
&=&\sigma _{_{\Sigma P}}\circ s_{_{M\Sigma}}\circ\pi _{_{PM}}(u)\nonumber\\
&=&\sigma _{_{\Sigma P}}\circ s_{_{M\Sigma}}(x)\nonumber\\
&=:&\sigma _\xi (x)\,,\label{sigmaxi03}
\end{eqnarray}
showing the coincidence of the images of sections $\sigma _{_{\Sigma P}}:\Sigma\rightarrow P$ and $\sigma _\xi :M\rightarrow P$. Accordingly, transformations (\ref{secttrans2}) can also be understood as
\begin{equation}
L_g\circ\sigma _{_{\Sigma P}}(x\,,\xi )=R_h\circ\sigma _{_{\Sigma P}}(x\,,\xi ')\,.\label{nonlintrans1}
\end{equation}
Depending on the alternative formulations (\ref{secttrans2}) and (\ref{nonlintrans1}), the interpretation of gauge transformations is twofold. When considering the group left action $L_g$ on sections $\sigma _\xi :M\rightarrow\Sigma\rightarrow P$ as in (\ref{secttrans2}), the verticality condition (\ref{gautrans02}) is satisfied as $\pi _{_{PM}}\circ L_g =\pi _{_{PM}}$, so that points $x\in M$ remain unchanged. Thus, (\ref{secttrans2}) is a vertical bundle automorphism assimilable to ordinary gauge transformations. Contrarily, in (\ref{nonlintrans1}) $L_g$ transforms sections $\sigma _{_{\Sigma P}}(x\,,\xi\,)$ of $P(\Sigma\,,H)$ into sections attached to different points $(x\,,\xi\,'\,)$ of $\Sigma$, while simultaneously displacing them vertically along $H$ fiber-branches by means of $R_h$. Obviously, the verticality condition (\ref{gautrans02}) does not hold with respect to the {\it intermediate base space} $\Sigma$ since $\pi _{_{P\Sigma}}\circ L_g\neq\pi _{_{P\Sigma}}$. Non-vertical transformations (\ref{nonlintrans1}) mapping $H$-fiber-branches to fiber branches defined on different $\Sigma$-points constitute the kind of action required by Lord for spacetime groups in which translations are present. Fig.2 illustrates both transformations (\ref{secttrans2}) and (\ref{nonlintrans1}), showing that $L_g$-related $H$-fibers $\pi _{_{P\Sigma}}^{-1}(x\,,\xi\,)$ and $\pi _{_{P\Sigma}}^{-1}(x\,,\xi\,'\,)$ of $P(\Sigma\,,H)$, attached to different points of the {\it intermediate space} $\Sigma$, are at the same time branches of a unique fiber $\pi _{_{PM}}^{-1}(x)$ of $P(M\,,G)$ over a single point $x\in M$.
\begin{figure}[h]
\begin{center}
\includegraphics[totalheight=63mm,width=89mm]{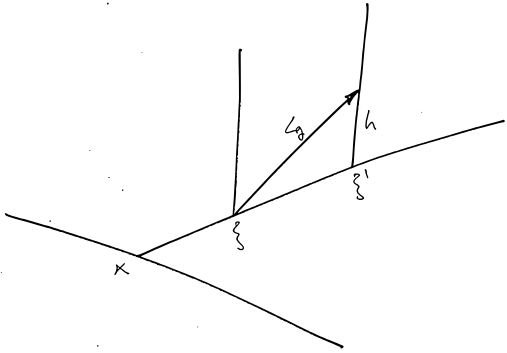}
\caption{\label{Fig2}Gauge transformations}
\end{center}
\end{figure}

Let us now show two different forms of gauge transformations of connections derived from the composite bundle formalism, coinciding respectively with those of linear and nonlinear realizations as can be found in Appendix A. Active gauge transformations of ordinary gauge potentials (\ref{pbconn03}) were discussed in \cite{Tresguerres:2002uh}. (We introduce a minor correction consisting in the change $g\rightsquigarrow g^{-1}$ for reasons of consistency with (\ref{nlr09}) and (\ref{nlr12}).) When considering $G$-group elements $g=e^\epsilon \approx 1+\epsilon$, where $\epsilon $ are infinitesimal $G$-algebra-valued parameters, gauge transformations induced by (\ref{secttrans1}) read
\begin{equation}
\delta A_{_M}= ( L_{g^{-1}}\circ\sigma _{_{MP}})^*\,\omega  -\sigma _{_{MP}}^*\,\omega \approx -(\,d\epsilon +[ A_{_M}\,,\epsilon\,])\,.\label{varlcon}
\end{equation}
In \cite{Tresguerres:2002uh} we also studied the transformation properties of modified gauge potentials (which we identify with nonlinear ones), defined as follows. Besides the pullback (\ref{pbconn02})-(\ref{pbconn03}) of the connection 1-form (\ref{connform1})-(\ref{connform2}) by zero sections of the bundle, let us also consider the pullback by a different section, namely
\begin{eqnarray}
\Gamma _{_M} &:=& s_{_{M\Sigma}}^*\,\Gamma _{_{\Sigma}}\nonumber\\
&=& s_{_{M\Sigma}}^*\,\sigma _{_{\Sigma P}}^*\,\omega\nonumber\\
&=& \sigma _\xi ^*\,\omega\,.\label{pbconn03bis}
\end{eqnarray}
(See (\ref{sigmaxi}).) The gauge transformations of (\ref{pbconn03bis}), in terms of $H$-group elements $h=e^\mu \approx 1+\mu$, with $\mu$ as $H$-algebra-valued parameters, were found to be
\begin{equation}
\delta\Gamma _{_M}= ( L_{g^{-1}}\circ\sigma _\xi )^*\,\omega  -\sigma _{\xi '}^*\,\omega \approx -(\,d\mu +[\Gamma _{_M}\,,\mu\,])\,.\label{varnlcon}
\end{equation}
Connections $\Gamma _{_M}$ take values on the Lie algebra of the whole group $G$. However, from (\ref{varnlcon}) one reads out that only
those of its components defined on the $H$-algebra transform inhomogeneously as true connections, while its components with values on the remaining algebra elements of $G/H$ transform as $H$-tensors. This is a feature which the present composite fiber bundle formalism has in common with nonlinear realizations \cite{Coleman:1969sm}--\cite{Tiemblo:2005sx}, thus providing a geometrical interpretation for them . Moreover, in composite bundles built from a principal bundle $P(M\,,G)$, the total symmetry remains that of the structure group $G$, while in the bundle sector $P(\Sigma\,,H)$ the subgroup $H\subset G$ becomes the explicit symmetry.

\section{Geometry of the Poincar\'e composite bundle with $H=$Lorentz}

Having established the main features of composite fiber bundles, we proceed to study the particular case of $G=$Poincar\'e with $H=$Lorentz and the resulting geometry. In the present Section we introduce a number of formal elements necessary to develop our program.

\subsection{Invariant 1-forms of the Poincar\'e group}

In order to build Poincar\'e bundles --composite or not-- and their tangent and cotangent bundle spaces, first we have to introduce left and right Poincar\'e invariant 1-forms and vectors. The Poincar\'e group is characterized by its Lorentz generators $\Lambda _{\alpha\beta}$ and translational generators $P_\alpha\hskip0.1cm (\alpha\,,\beta =\,0\,,...\,,3)\,$, satisfying the commutation relations
\begin{eqnarray}
\left[\Lambda _{\alpha\beta }\,,\Lambda _{\mu\nu }\right] &=& -i\,\left( o_{\alpha [\mu } \Lambda _{\nu ]\beta} - o_{\beta [\mu }\Lambda _{\nu ]\alpha }\right)\,,\label{Lor}\\
\left[\Lambda _{\alpha\beta }\,, P_\mu\,\right] &=& \hskip0.20cm i\,o_{\mu [\alpha }P_{\beta ]}\,,\label{mix}\\
\left[ P_\alpha\,, P_\beta\,\right] &=& \hskip0.20cm0\,,\label{trans}
\end{eqnarray}
with the Minkowski metric $o_{\alpha\beta}:=\,diag(-\,+\,+\,+\,)$. According to (\ref{condit1}), we parametrize arbitrary group elements $\tilde{g}\in G=$Poincar\'e as
\begin{equation}
\tilde{g} =b\cdot a\,,\quad{\rm with}\quad a= e^{i\,\lambda ^{\alpha\beta}\Lambda _{\alpha\beta}}\,,\quad b = e^{-i\,\xi ^\mu P_\mu}\,,\label{poinc01}
\end{equation}
being $a\in H=$Lorentz and $b\in G/H =$Translations. In terms of (\ref{poinc01}), we calculate the left invariant Maurer-Cartan form as
\begin{eqnarray}
\Theta _{_G} &:=&\,\tilde{g}^{-1} d\tilde{g}\nonumber\\
&=&\,a^{-1} da + a^{-1}( b^{-1}db\,)\,a\nonumber\\
&=& \,\Theta _{^{(\Lambda )}}^{\alpha\beta }\Lambda _{\alpha\beta} +\Theta _{^{(P)}}^\mu P_\mu\,,\label{poinc02}
\end{eqnarray}
with
\begin{eqnarray}
\Theta _{^{(\Lambda )}}^{\alpha\beta } &:=& \,i\,u^{\,\lambda\alpha} du_\lambda {}^\beta\,,\label{poinc03}\\
\Theta _{^{(P)}}^\mu &:=& -i\,d\xi ^\lambda u_\lambda{}^\mu\,,\label{poinc03bis}
\end{eqnarray}
where we defined the Lorentz matrices
\begin{equation}
u_\alpha{}^\beta :=(e^\lambda\,) _\alpha{}^\beta :=\delta _\alpha ^\beta +\lambda _\alpha{}^\beta +{1\over{2!}}\lambda _\alpha{}^\gamma \lambda _\gamma{}^\beta +...\,,\label{poinc04}
\end{equation}
such that $\left( u^{-1}\right) _\alpha{}^\beta =u^{\,\beta}{}_\alpha\,$. In parallel to (\ref{poinc02}), we introduce the right invariant forms
\begin{eqnarray}
\overline{\Theta}_{_G} &:=&\,d\tilde{g}\cdot\tilde{g}^{-1}\nonumber\\
&=&\,da\cdot a^{-1} +\left[\,d(ba)\cdot (ba)^{-1} - da\cdot a^{-1}\,\right]\nonumber\\
&=&\,\overline{\Theta}_{^{(\Lambda )}}^{\alpha\beta }\Lambda _{\alpha\beta} +\overline{\Theta}_{^{(P)}}^\mu P_\mu\,,\label{poinc09}
\end{eqnarray}
being
\begin{eqnarray}
\overline{\Theta}_{^{(\Lambda )}}^{\alpha\beta } &:=&\,i\,du^{\alpha\lambda}\,u^{\,\beta}{}_\lambda\,,\label{poinc10}\\
\overline{\Theta}_{^{(P)}}^\mu &:=& -i\,d\left(\xi ^\lambda u_\lambda{}^\nu\,\right) u^{\,\mu}{}_\nu =-i\,d\xi ^\mu -\overline{\Theta}_{^{(\Lambda )}}^{\nu\mu}\xi _\nu\,.\label{poinc11}
\end{eqnarray}
Besides these quantities, a particular representation of the Poincar\'e group will also play a role in the following.

\subsection{5-D representation of the Poincar\'e group}

The Poincar\'e group generators in (\ref{Lor})--(\ref{trans}) admit the 5-dimensional matrix representation
\begin{eqnarray}
(\Lambda _{\alpha\beta} )_A{}^B &=&-i\,o_{A[\alpha}\,\delta _{\beta ]}^B\,,\label{5dpoinc01}\\
(P_\mu )_A{}^B &=& -i\,l^{-1}\,\delta _A^5\,\delta _\mu ^B\,,\label{5dpoinc02}
\end{eqnarray}
with $\alpha\,,\beta$ running from $0$ to $3$, and $A\,,B =0,1,2,3,5\,$, and with $l$ as a dimensional constant. The object $o_{AB}$ in (\ref{5dpoinc01}) is a symmetric matrix whose components
\begin{equation}
o_{\alpha\beta} = {\rm diag}(-+++)\label{5dpoinc03}
\end{equation}
are identical with the Minkowski metric, and
\begin{equation}
o_{\alpha 5}=0\,,\label{5dpoinc04}
\end{equation}
while the components $o_{55}$, being undetermined, can be fixed at convenience. Using (\ref{5dpoinc01}) and (\ref{5dpoinc02}), it is possible to construct a 5-dimensional representation of Poincar\'e group elements decomposed as in (\ref{poinc01}). To do so, we calculate
\begin{equation}
a_A{}^B := ( e^{i\,\lambda ^{\alpha\beta}\Lambda _{\alpha\beta}})_A{}^B = \delta _A^5\delta _5^B + u^{\alpha\beta}\,o_{A\alpha}\delta _\beta ^B\,,\label{5dpoinc05}
\end{equation}
where $u_\alpha{}^\beta $ is given by (\ref{poinc04}), and
\begin{eqnarray}
b_A{}^B &:=& ( e^{-i\,\xi ^\mu P_\mu})_A{}^B\nonumber\\
&=& \delta _A^B -i\,\xi ^\mu\,(P_\mu )_A{}^B\nonumber\\
&=& \delta _A^B -l^{-1}\,\xi ^\mu\,\delta _A^5\,\delta _\mu ^B\,,\label{5dpoinc06}
\end{eqnarray}
and, as in (\ref{poinc01}), we find their product
\begin{eqnarray}
\tilde{g}_A{}^B &=& b_A{}^C\,a_C{}^B\nonumber\\
&=&\delta _A^5\delta _5^B + u^{\alpha\beta}\,o_{A\alpha}\delta _\beta ^B -l^{-1}\,\xi ^\mu u_\mu{}^\beta\,\delta _A^5\,\delta _\beta ^B\,,\label{5dpoinc07}
\end{eqnarray}
whose $5\times 5$ components read
\begin{equation}
\tilde{g}_A{}^B =
\begin{pmatrix}
\tilde{g}_\alpha{}^\gamma & \tilde{g}_\alpha{}^5\\
\tilde{g}_5{}^\gamma & \tilde{g}_5{}^5
\end{pmatrix}
=\begin{pmatrix}
u_\alpha{}^\gamma & 0\\
-l^{-1}\,\xi ^\mu u_\mu{}^\gamma & 1
\end{pmatrix}
\,.\label{5dpoinc08}
\end{equation}
For later convenience we also determine the inverse group element
\begin{eqnarray}
(\tilde{g}^{-1})_A{}^B &=& (a^{-1})_A{}^C\,(b^{-1})_C{}^B\nonumber\\
&=&\delta _A^5\delta _5^B + (u^{-1})^{\alpha\beta}\,o_{A\alpha}\delta _\beta ^B +l^{-1}\,\xi ^\mu\,\delta _A^5\,\delta _\mu ^B\,,\nonumber\\
\label{5dpoinc09}
\end{eqnarray}
with $5\times 5$ matrix components
\begin{equation}
(\tilde{g}^{-1})_A{}^B =
\begin{pmatrix}
(\tilde{g}^{-1})_\gamma{}^\beta & (\tilde{g}^{-1})_\gamma{}^5\\
(\tilde{g}^{-1})_5{}^\beta & (\tilde{g}^{-1})_5{}^5
\end{pmatrix}
=\begin{pmatrix}
(u^{-1})_\gamma{}^\beta & 0\\
l^{-1}\,\xi ^\beta & 1
\end{pmatrix}
\,.\label{5dpoinc10}
\end{equation}
The usefullness of all these quantities will become apparent later.

\subsection{Invariant Poincar\'e vectors}

Left invariant vectors dual to (\ref{poinc03}), (\ref{poinc03bis}) are respectively given by
\begin{eqnarray}
L^{_{(\Lambda )}}_{\alpha\beta }&:=&-(\Lambda _{\alpha\beta} )_M{}^N\,(\tilde{g}^{-1})_N{}^L\,{{\partial}\over{\partial (\tilde{g}^{-1})_M{}^L}}\nonumber\\
&=& -i\,u_{\,\lambda [\,\alpha}{\partial\over{\partial u _\lambda{}^{\,\beta ]}}}\,,\label{poinc05}\\
L^{_{(P)}}_\mu &:=&-(P_\mu\,)_M{}^N\,(\tilde{g}^{-1})_N{}^L\,{{\partial}\over{\partial (\tilde{g}^{-1})_M{}^L}}\nonumber\\
&=& \,i\,u^\nu{}_\mu {\partial\over{\partial\xi ^{\,\nu}}}\,,\label{poinc06}
\end{eqnarray}
related to (\ref{poinc02}) as
\begin{eqnarray}
L^{_{(\Lambda )}}_{\alpha\beta }\rfloor\Theta _{_G} &=&\,\Lambda _{\alpha\beta}\,,\label{poinc07}\\
L^{_{(P)}}_\mu\rfloor\Theta _{_G} &=&\,P_\mu\,,\label{poinc08}
\end{eqnarray}
and satisfying commutation relations formally identical to (\ref{Lor})-(\ref{trans}). On the other hand, the dual right invariant vectors of (\ref{poinc09})--(\ref{poinc11}) read
\begin{eqnarray}
\overline{L}^{_{(\Lambda )}}_{\alpha\beta } &:=&-(\tilde{g}^{-1})_M{}^N\,(\Lambda _{\alpha\beta} )_N{}^L\,{{\partial}\over{\partial (\tilde{g}^{-1})_M{}^L}}\nonumber\\
&=& \,i\left(u_{[\,\alpha}{}^\lambda {\partial\over{\partial u^{\,\beta ]\lambda}}} +\xi _{[\,\alpha}{\partial\over{\partial\xi ^{\,\beta ]}}}\right) \,,\label{poinc12}\\
\overline{L}^{_{(P)}}_\mu &:=&-(\tilde{g}^{-1})_M{}^N\,(P_\mu\,)_N{}^L\,{{\partial}\over{\partial (\tilde{g}^{-1})_M{}^L}}\nonumber\\
&=& \,i\,{\partial\over{\partial\xi ^{\,\mu}}}\,.\label{poinc13}
\end{eqnarray}
In (\ref{poinc12})-(\ref{poinc13}) one recognizes the angular momentum and the linear momentum generators respectively \cite{Tresguerres:2002uh}, being the total angular momentum (\ref{poinc12}) decomposed into an intrinsic and an orbital piece as
\begin{equation}
\overline{L}^{_{(\Lambda )}}_{\alpha\beta }=\overline{L}^{_{(Int)}}_{\alpha\beta }+\overline{L}^{_{(Orb)}}_{\alpha\beta }\,,\qquad\overline{L}^{_{(Orb)}}_{\alpha\beta }:=\,i\,\xi _{[\,\alpha}{\partial\over{\partial\xi ^{\,\beta ]}}}\,.\label{poinc14}
\end{equation}
Vectors (\ref{poinc12})-(\ref{poinc14}) with (\ref{poinc09}) satisfy
\begin{eqnarray}
\overline{L}^{_{(Int)}}_{\alpha\beta }\rfloor\overline{\Theta}_{_G} &=&\,\Lambda _{\alpha\beta} -\xi_{[\,\alpha}P_{\beta ]}\,,\label{poinc15}\\
\overline{L}^{_{(Orb)}}_{\alpha\beta }\rfloor\overline{\Theta}_{_G} &=&\,\xi_{[\,\alpha}P_{\beta ]}\,,\label{poinc16}\\
\overline{L}^{_{(P)}}_\mu\rfloor\overline{\Theta}_{_G} &=&\,P_\mu\,.\label{poinc17}
\end{eqnarray}
The technical results established in the present Section III provide the support for what follows.

\section{Connections on the Poincar\'e composite bundle}

Let us particularize definitions (\ref{connform1}), (\ref{connform2}) of the connection 1-form for $G=$the Poincar\'e group, with $H=$Lorentz and $G/H =$Translations. Ordinary (linear) gauge potentials (\ref{pbconn03}) of the Poincar\'e group are of the form
\begin{equation}
A_{_M} =\sigma _{_{MP}}^*\,\omega = -i(\,dx^i {\buildrel (T)\over{\Gamma _i\,^\mu}} P_\mu + dx^i \Gamma _i^{\alpha\beta}\Lambda _{\alpha\beta }\,)\,,\label{pbconn04}
\end{equation}
including standard translational and Lorentz potentials. Replacing (\ref{pbconn04}) in (\ref{connform2}), and taking into account the value of $b$ given in (\ref{poinc01}), the $T^*(\Sigma )$ quantity (\ref{pbconn01}) becomes
\begin{equation}
\Gamma _{_{\Sigma}}= \sigma _{_{\Sigma P}}^*\,\omega = -i\,(\vartheta _{_\Sigma}^\mu P_\mu + \pi _{_{\Sigma M}}^*\,\Gamma ^{\alpha\beta}\Lambda _{\alpha\beta }\,)\,,\label{poincconnform3}
\end{equation}
with a Lorentz contribution $\pi _{_{\Sigma M}}^*\,\Gamma ^{\alpha\beta}$ built from
\begin{equation}
\Gamma ^{\alpha\beta} = dx^i\,\Gamma _i^{\alpha\beta}\,,\label{connform5}
\end{equation}
as appearing in (\ref{pbconn04}), and including the modified translational connection form defined as
\begin{equation}
\vartheta _{_{\Sigma}}^\mu := d_{_{\Sigma}}\xi ^\mu + \pi _{_{\Sigma M}}^* dx^i (\Gamma _{i\nu}{}^\mu \xi\,^\nu +{\buildrel (T)\over{\Gamma _i\,^\mu}}\,)\,,\label{connform4}
\end{equation}
closely related to that of NLR's \cite{Tresguerres:2002uh} \cite{Tiemblo:2005js}, which will play a central role in the following. Observe that, from (\ref{poincconnform3}), we can either recover (\ref{pbconn04}) according to (\ref{pbconn02}), or we can get (\ref{pbconn03bis}) pulling back (\ref{poincconnform3}) to $M$ by means of $s_{_{M\Sigma}}^*$, yielding the Poincar\'e nonlinear connection defined on the base space $M$, namely
\begin{eqnarray}
\Gamma _{_M} = \sigma _\xi ^*\,\omega &=& -i\,(\,dx^i e_i{}^\mu P_\mu + dx^i\Gamma _i ^{\alpha\beta}\Lambda _{\alpha\beta }\,)\nonumber\\
&=& -i\,(\,\vartheta _{_M}^\mu P_\mu + \Gamma ^{\alpha\beta}\Lambda _{\alpha\beta }\,)\,,\label{nlinconn2}
\end{eqnarray}
where, instead of the ordinary translational connection $dx^i {\buildrel (T)\over{\Gamma _i\,^\mu}}$ present in (\ref{pbconn04}), now we get the nonlinear translational connection
\begin{equation}
\vartheta _{_M}^\mu =s_{_{M\Sigma}}^*\vartheta _{_{\Sigma}}^\mu := dx^i e_i{}^\mu\,,\label{variety05}
\end{equation}
(pulled back to $M$), which we identify with ordinary tetrads \cite{Tresguerres:2002uh} provided with an internal structure derived from (\ref{connform4}) and (\ref{variety05}), manifesting itself in the {\it vierbeins} $e_i{}^\mu\,$, which take the form
\begin{equation}
e_i{}^\mu := \partial _i\xi ^\mu +\Gamma _{i\nu}{}^\mu\xi\,^\nu +{\buildrel (T)\over{\Gamma _i\,^\mu}}\,.\label{variety06}
\end{equation}
Putting together (\ref{variety05}) and (\ref{variety06}), the tetrads on $M$ can be written as
\begin{equation}
\vartheta _{_M}^\mu = D\xi ^\mu +{\buildrel (T)\over{\Gamma ^\mu}}\,,\label{variety06bis}
\end{equation}
where we introduce the notation of the covariant derivative of $\xi ^\mu $ \cite{Tresguerres:2002uh} \cite{Tiemblo:2005js}.

Let us now make use of (\ref{poincconnform3}) and of the value of $a$ given in (\ref{poinc01}) to determine the form of the Poincar\'e connection (\ref{connform1}), which for later convenience we decompose into two connection pieces as
\begin{equation}
\omega = {\buildrel (P)\over{\omega }} +{\buildrel (\Lambda )\over{\omega}}\,,\label{variety18}
\end{equation}
corresponding to different orientations along the Lie algebra bases of $G/H$ (translations) and $H$ (Lorentz group) respectively. We find
\begin{equation}
{\buildrel (P)\over{\omega }} = a^{-1}( -i\pi _{_{P\Sigma}}^*\vartheta _{_{\Sigma}}^\mu P_\mu\,)\,a =-i\,\pi _{_{P\Sigma}}^*\vartheta _{_{\Sigma}}^\mu\, u_\mu{}^\nu P_\nu \label{variety19}
\end{equation}
as the part of the total connection relevant for the fibred space $\Sigma\rightarrow M$, and
\begin{eqnarray}
{\buildrel (\Lambda )\over{\omega}} &=& a^{-1} ( d_{_P} -i\,\pi _{_{PM}}^*\Gamma ^{\alpha\beta}\Lambda _{\alpha\beta}\,)\,a\nonumber\\
&=&(\,\overline{\Theta}_{^{(\Lambda )}}^{\alpha\beta } -i\,\pi _{_{PM}}^*\Gamma ^{\alpha\beta}\,)\,u_\alpha{}^\mu u_\beta{}^\nu \Lambda _{\mu\nu}\,,\label{variety20}
\end{eqnarray}
with definitions (\ref{poinc10}) and (\ref{connform5}), as the part of the connection corresponding to the bundle $P\rightarrow\Sigma$.

\section{Associated 5-dimensional vector bundle }

\subsection{The 5-D vector basis}

Group elements $\tilde{g}\in G$ in the 5-dimensional matrix representation (\ref{5dpoinc07}) are suitable to act as invertible linear operators transforming a 5-dimensional vector space $V$ to itself. The representation space $V$ can be identified with $\mathbb{R}^5$ provided it is equipped with an appropriate basis. Let us choose the natural vector basis $\{\,{\buildrel o\over{e}}_A\,\}$ with components $({\buildrel o\over{e}}_A)^B =\delta _A^B$.

Having introduced this 5-dimensional representation of the Poincar\'e group, next we construct for it the $P$-associated vector bundle. The corresponding basis vectors $e_A (x)$ are defined to be equivalent classes $\left|\left(\,\sigma _{_{MP}}(x)\,,{\buildrel o\over{e}}_A\,\right)\right|$ belonging to $P\times _G \mathbb{R}^5$, where an equivalence relation exists between different representatives, consisting in the simultaneous $G$-transformation of both elements of the ordered pair (with $G$ acting on the $P$-section $\sigma _{_{MP}}$ on the right and on the basis vectors ${\buildrel o\over{e}}_A$ on the left) \cite{Nakahara}. That is
\begin{eqnarray}
e_A (x) &:=& \left|\left(\,\sigma _{_{MP}}(x)\,,{\buildrel o\over{e}}_A\,\right)\right|\label{5dpoinc11}\\
&=& \left|\left(\,R_{\tilde{g}}\circ\sigma _{_{MP}}(x)\,,(\tilde{g}^{-1})_A{}^B\,{\buildrel o\over{e}}_B\,\right)\right|\,,\label{5dpoinc12}
\end{eqnarray}
which can be displayed as
\begin{equation}
e_A (x) = \binom{e_\alpha (x)}{e_5 (x)}\label{5dpoinc13}
\end{equation}
in order to single out the fifth basis vector component. In principle, no obvious spacetime interpretation seems to exist for such 5-dimensional vector basis in a four-dimensional world. However, the role of $e_5 (x)$ as the frame origin necessary to complete an affine frame --in rigor, a Poincar\'e frame-- becomes apparent by introducing, in addition to the vector space associated to the principal bundle $P\rightarrow M$, an analogous basis on the bundle sector $P\rightarrow\Sigma$. The modified fifth component will clarify the role played by the original $e_5 (x)$.

To build the second vector basis, referred to the intermediate space $\Sigma$, we proceed as follows. First we realize that, according to the equivalence relation defining the vector space associated to the total bundle $P(M\,,G)$, the following chain of identities for ordered pairs belonging to $P\times\mathbb{R}^5$ holds
\begin{eqnarray}
\left(\sigma _{_{MP}}(x)\,,{\buildrel o\over{e}}_A\right) &\simeq & \left( R_{\tilde{g}}\circ\sigma _{_{MP}}(x)\,,(\tilde{g}^{-1})_A{}^B\,{\buildrel o\over{e}}_B\right)\nonumber\\
&=& \left( R_a\circ R_b\circ\sigma _{_{MP}}(x)\,,( a^{-1})_A{}^B ( b^{-1})_B{}^C\,{\buildrel o\over{e}}_C\right)\nonumber\\
&=& \left( R_a\circ\sigma _\xi (x)\,,( a^{-1})_A{}^B\,{\buildrel o\over{\widehat{e}}}_{\,B}\right)\nonumber\\
&\simeq & \left( \sigma _\xi (x)\,,{\buildrel o\over{\widehat{e}}}_{\,A}\right)\,,\label{5dpoinc14}
\end{eqnarray}
where we made use of (\ref{condit1}), (\ref{sigmaxibis}) and (\ref{5dpoinc09}) and we introduced the redefined natural vector basis
\begin{equation}
{\buildrel o\over{\widehat{e}}}_A := (b^{-1})_A{}^B\,{\buildrel o\over{e}}_B\label{5dpoinc15}
\end{equation}
in terms of the inverse matrix of (\ref{5dpoinc06}). All quantities in (\ref{5dpoinc14}) constitute different representatives of the equivalence class (\ref{5dpoinc11}).

However, by replacing in the last expression of (\ref{5dpoinc14}) the section $\sigma _\xi (x)$ with $\sigma _{_{\Sigma P}}( x\,,\xi )$, as suggested by the coincidence of their respective images guaranteed by (\ref{sigmaxi03}), we get the vector basis of a different --although closely related-- vector bundle associated to the principal bundle sector $P(\Sigma\,,H)$. So, in analogy to (\ref{5dpoinc11}), defined on $M$, and having at sight (\ref{5dpoinc14}), we define on $\Sigma$ the vector basis
\begin{eqnarray}
\widehat{e}_A (x\,,\xi\,) &:=& \left|\left(\,\sigma _{_{\Sigma P}}( x\,,\xi )\,,{\buildrel o\over{\widehat{e}}}_{\,A}\,\right)\right|\label{5dpoinc16}\\
\emph{\emph{}}&=& \left|\left(\,R_a\circ\sigma _{_{\Sigma P}}( x\,,\xi )\,,( a^{-1})_A{}^B\,{\buildrel o\over{\widehat{e}}}_{\,B}\,\right)\right|\,.\label{5dpoinc17}
\end{eqnarray}
Notice that the equivalence relation, according to (\ref{5dpoinc17}), is defined exclusively in terms of $a\in H$, not involving $b\in G/H$, so that $\widehat{e}_A (x\,,\xi\,'\,)\neq\widehat{e}_A (x\,,\xi\,)$. A relevant difference of (\ref{5dpoinc16}) with respect to (\ref{5dpoinc11}) consists in that, contrarily to (\ref{5dpoinc11}), depending on the natural basis ${\buildrel o\over{e}}_{\,A}$, the vectors (\ref{5dpoinc16}) are built with ${\buildrel o\over{\widehat{e}}}_{\,A}$, defined by (\ref{5dpoinc15}). To make explicit the role played by ${\buildrel o\over{\widehat{e}}}_{\,A}$ (with hat), let us define the (\ref{5dpoinc16})-analogous quantity
\begin{equation}
e_A (x\,,\xi\,) := \left|\left(\,\sigma _{_{\Sigma P}}( x\,,\xi )\,,{\buildrel o\over{e}}_{\,A}\,\right)\right|\,,\label{5dpoinc16bis}
\end{equation}
differing from (\ref{5dpoinc16}) merely in that, in it, (\ref{5dpoinc15}) is replaced by the natural basis ${\buildrel o\over{e}}_{\,A}$ (without hat). Then, in terms of (\ref{5dpoinc16bis}) (and of the inverse of (\ref{5dpoinc06})) we can rewrite (\ref{5dpoinc16}) as
\begin{eqnarray}
\widehat{e}_A (x\,,\xi\,) &=& (b^{-1})_A{}^B\,e_B (x\,,\xi\,)\nonumber\\
&=& \left[\,\delta _A^B +i\,\xi ^\mu (P_\mu )_A{}^B\,\right]\,e_B (x\,,\xi\,)\nonumber\\
&=& e_A (x\,,\xi\,)+l^{-1}\,\delta _A^5\,\xi ^\mu\,e_\mu (x\,,\xi\,)\,.\label{5dpoinc18}
\end{eqnarray}
We point out the non-casual formal analogy existing between (\ref{5dpoinc18}) and the representation fields (\ref{nlr06}) of NLR's. In parallel to (\ref{5dpoinc13}), eq.(\ref{5dpoinc18}) can be expressed as
\begin{equation}
\widehat{e}_A (x\,,\xi\,) = \binom{e_\alpha (x\,,\xi\,)}{l^{-1}\,[\,l\,e_5 (x\,,\xi\,) +\xi ^\mu\,e_\mu (x\,,\xi\,) ]}\,.\label{5dpoinc19}
\end{equation}
Making abstraction of the different base spaces --$M$ and $\Sigma$ respectively-- where each vector basis is defined, the main new feature exhibited by (\ref{5dpoinc19}) as compared with (\ref{5dpoinc13}) consists in the modified structure of its fifth vector component
\begin{equation}
\widehat{e}_5 = l^{-1}\,(\,l\,e_5 +\xi ^\mu\,e_\mu )\,.\label{5dpoinc20}
\end{equation}
This quantity admits at least two different interpretations. According to one of them, (\ref{5dpoinc20}) constitutes a 5-dimensional vector $\widehat{e}_5 = l^{-1}\,\xi ^A \,e_A$ with the constant $l$ as its fifth component. The second possible interpretation, which we embrace hereafter, regards both, $l \widehat{e}_5$ and $l e_5$, as points joined by the fourvector-like quantity $\xi ^\mu\,e_\mu $. Accordingly, we choose one of these points as an origin
\begin{equation}
\mathfrak{o} := l\,e_5\,,\label{5dpoinc21}
\end{equation}
and we regard the other one as an object formalizing position. Let us denote it as
\begin{equation}
p := l\,\hat{e}_5\,.\label{5dpoinc22}
\end{equation}
Replacing (\ref{5dpoinc21}) and (\ref{5dpoinc22}) in (\ref{5dpoinc20}) we get
\begin{equation}
p = \mathfrak{o} +\xi ^\mu\,e_\mu\,,\label{5dpoinc23}
\end{equation}
showing, in fact, the aptitude of (\ref{5dpoinc22}) to depict positions relatively to the origin (\ref{5dpoinc21}) and measured by the {\it position vector} $\xi ^\mu\,e_\mu$. The translational parameters $\xi ^\mu $ introduced through $b$ in (\ref{poinc01}) (or through $b_A{}^B$ in (\ref{5dpoinc06})) acquire in a natural way the character of position vector components  --as constitutive elements of the gauge formalization (\ref{5dpoinc23}) of position--. Later we will show that these fields $\xi ^\mu $ play a central role in the description of motion, in close analogy to the spacetime coordinates of Special Relativity. We will check below that our interpretation is supported by the gauge transformation properties of the different pieces involved in (\ref{5dpoinc23}); see (\ref{5dpoinc74})--(\ref{5dpoinc77}).

With the notation (and interpretation) (\ref{5dpoinc23}) for the fifth component, the vector basis (\ref{5dpoinc19}) takes the form
\begin{equation}
\widehat{e}_A (x\,,\xi\,) = \binom{e_\alpha (x\,,\xi\,)}{l^{-1}\,p\,(x\,,\xi\,)\,}\,,\label{5dpoinc24}
\end{equation}
which presupposes that we also interpret the fifth component of (\ref{5dpoinc16bis}) as an origin (\ref{5dpoinc21}), so that the 5-D Poincar\'e basis (\ref{5dpoinc16bis}) --and its analogous (\ref{5dpoinc13})-- is to be seen as a sort of affine frame.

\subsection{Gauge transformations of 5-D basis vectors}

The geometrical as much as the physical meaning of (\ref{5dpoinc13}) and (\ref{5dpoinc24}) is further clarified by examining their respective responses to two kinds of action on them. In the present paragraph we examine their gauge transformations, making use of the notation and the results of Appendix B. Changes induced by a different manipulation (lateral displacement) of these objects will be considered later.

Gauge variations of (\ref{5dpoinc13}) (or (\ref{5dpoinc11})) result from the application of a gauge transformation (\ref{secttrans1}) to the section defined in (\ref{5dpoinc11}). Even though (\ref{secttrans1}) involves the left action $L_g$ (with $g=g(x)$) rather than the right one $R_g$ with which we know how to operate inside (\ref{5dpoinc11}), we assume, as in Appendix B, that $\sigma _{_{MP}}^{-1}(x)\cdot g(x)\cdot\sigma _{_{MP}}(x) =g(x)$, so that $L_g\circ\sigma _{_{MP}} = R_g\circ\sigma _{_{MP}}$. Then, by definition, the gauge transformed of (\ref{5dpoinc11}) reads
\begin{eqnarray}
e'_A (x) &:=& \left|\left(\,L_g\circ\sigma _{_{MP}}(x)\,,{\buildrel o\over{e}}_A\,\right)\right|\nonumber\\
&=& \left|\left(\,R_g\circ\sigma _{_{MP}}(x)\,,{\buildrel o\over{e}}_A\,\right)\right|\nonumber\\
&=& \left|\left(\,\sigma _{_{MP}}(x)\,,g_A{}^B\,{\buildrel o\over{e}}_B\,\right)\right|\nonumber\\
&=& g_A{}^B\,e_{_B} (x)\,.\label{5dpoinc25}
\end{eqnarray}
In particular we take $g_A{}^B$ to be the 5-dimensional representation of the infinitesimal group element (\ref{formula02}), that is
\begin{eqnarray}
g_A{}^B  =&& (\,e^{i\,\beta ^{\alpha\beta}\Lambda _{\alpha\beta}}\,e^{i\,\epsilon ^\mu P_\mu}\,)_A{}^B\nonumber\\
\approx &&\delta _A^B + i\,\beta ^{\alpha\beta}(\Lambda _{\alpha\beta} )_A{}^B +i\,\epsilon ^\mu (P_\mu )_A{}^B\,.\label{5dpoinc26}
\end{eqnarray}
When replaced in (\ref{5dpoinc25}) using (\ref{5dpoinc01}) and (\ref{5dpoinc02}), it yields
\begin{equation}
e'_A (x) =e_A (x) + o_{A\alpha}\,\beta ^{\alpha\beta}\,e_\beta (x) + l^{-1}\,\delta _A^5\,\epsilon ^\mu\, e_\mu (x)\,,\label{5dpoinc27}
\end{equation}
so that the basis vector variations turn out to be
\begin{eqnarray}
\delta\,e_\alpha (x) &:=& e'_\alpha (x) -e_\alpha (x) =\beta _\alpha{}^\beta\,e_\beta (x)\,,\label{5dpoinc28}\\
\delta\,e_5 (x) &:=& e'_5 (x) - e_5 (x) = l^{-1}\epsilon ^\mu\,e_\mu (x)\,.\label{5dpoinc29}
\end{eqnarray}
Although not explicitly displayed, being $g=g(x)$ (more details in \cite{Tresguerres:2002uh}), the gauge parameters are understood to depend on coordinates $x\in M$, that is, $\beta _\alpha{}^\beta =\beta _\alpha{}^\beta (x)$ and $\epsilon ^\mu =\epsilon ^\mu (x)$.

On the other hand, gauge transformations of (\ref{5dpoinc24}) (or (\ref{5dpoinc16})) require a somewhat different treatment. The action of $L_g$ is in this case (\ref{nonlintrans1}), moving sections from $( x\,,\xi )\in\Sigma $ to $( x\,,\xi\,' )\in\Sigma $. Nevertheless, (\ref{nonlintrans1}) also induces standard {\it vertical} gauge transformations on $\Sigma $ along fiber branches $\pi ^{-1}( x\,,\xi\,' )$ as follows. Acting simultaneously on both ordered-pair elements of (\ref{5dpoinc16}), and denoting by ${\buildrel o\over{\widehat{e}\,'}}_A $ the transformed quantity
\begin{equation}
{\buildrel o\over{\widehat{e}\,'}}_A := (b'^{-1})_A{}^B\,{\buildrel o\over{e}}_B\,,\label{5dpoinc30}
\end{equation}
as defined by (\ref{5dpoinc15}), involving the (primed) (\ref{5dpoinc06})-inverse
\begin{equation}
(b'^{-1})_A{}^B =( e^{i\,\xi\,'^\mu P_\mu})_A{}^B =\delta _A^B +l^{-1}\,\xi\,'^\mu\,\delta _A^5\,\delta _\mu ^B\,,\label{5dpoinc31}
\end{equation}
we get
\begin{eqnarray}
\widehat{e}\,'_A (x\,,\xi\,'\,) &:=& \left|\left(\,L_g\circ\sigma _{_{\Sigma P}}( x\,,\xi )\,,{\buildrel o\over{\widehat{e}\,'}}_{\,A}\,\right)\right|\nonumber\\
&=& \left|\left(\,R_h\circ\sigma _{_{\Sigma P}}( x\,,\xi\,' )\,,{\buildrel o\over{\widehat{e}\,'}}_{\,A}\,\right)\right|\nonumber\\
&=& \left|\left(\,\sigma _{_{\Sigma P}}( x\,,\xi\,' )\,,h_A{}^B\,{\buildrel o\over{\widehat{e}\,'}}_{\,B}\,\right)\right|\nonumber\\
&=& h_A{}^B\,\widehat{e}_B (x\,,\xi\,'\,)\,.\label{5dpoinc32}
\end{eqnarray}
Replacing in (\ref{5dpoinc32}) the infinitesimal $H$-subgroup element (\ref{formula03}) in its 5-dimensional representation
\begin{equation}
h_A{}^B  = (\,e^{i\,\mu ^{\alpha\beta}\Lambda _{\alpha\beta}}\,)_A{}^B \approx \delta _A^B + i\,\mu ^{\alpha\beta}(\Lambda _{\alpha\beta} )_A{}^B\,,\label{5dpoinc33}
\end{equation}
with $\mu ^{\alpha\beta}=\beta ^{\alpha\beta}$ as derived in (\ref{5dpoinc41}), and with (\ref{5dpoinc01}), we find
\begin{equation}
\widehat{e}\,'_A (x\,,\xi\,'\,) =\widehat{e}_A (x\,,\xi\,'\,) + o_{A\alpha}\,\beta ^{\alpha\beta}\,\widehat{e}_\beta (x\,,\xi\,'\,)\,.\label{5dpoinc34}
\end{equation}
From (\ref{5dpoinc34}) we define the gauge variation
\begin{eqnarray}
\delta\,\widehat{e}_A (x\,,\xi\,'\,) &:=& \widehat{e}\,'_A (x\,,\xi\,'\,) -\widehat{e}_A (x\,,\xi\,'\,)\nonumber\\
&=& o_{A\alpha}\,\beta ^{\alpha\beta}\,\widehat{e}_\beta (x\,,\xi\,'\,)\,,\label{5dpoinc35}
\end{eqnarray}
constituting the vertical change of (\ref{5dpoinc16}) (over a transformed point $(x\,,\xi\,'\,)\in\Sigma\,$). Of course, the total transformation induced by $L_g$ on the basis vectors is
\begin{eqnarray}
\Delta\,\widehat{e}_A &:=& \widehat{e}\,'_A (x\,,\xi\,'\,) -\widehat{e}_A (x\,,\xi\,)\nonumber\\
&=&\delta\,\widehat{e}_A (x\,,\xi\,'\,) + \left[\,\widehat{e}_A (x\,,\xi\,'\,) -\widehat{e}_A (x\,,\xi\,)\,\right]\,.\nonumber\\
\label{5dpoinc36}
\end{eqnarray}
However, in the present context, the additional contribution in (\ref{5dpoinc36}) is to be seen as an induced (active) coordinate transformation of the base space $\Sigma $ (in analogy to Fig.2). More explicitly, gauge variations (\ref{5dpoinc35}) read
\begin{eqnarray}
\delta\,\widehat{e}_\alpha (x\,,\xi\,'\,) &:=& \widehat{e}\,'_\alpha (x\,,\xi\,'\,) -\widehat{e}_\alpha (x\,,\xi\,'\,)\nonumber\\
&=& \beta _\alpha{}^\beta\,\widehat{e}_\beta (x\,,\xi\,'\,)\,,\label{5dpoinc37}\\
\delta\,\widehat{e}_5 (x\,,\xi\,'\,) &:=& \widehat{e}\,'_5 (x\,,\xi\,'\,) -\widehat{e}_5 (x\,,\xi\,'\,) =0\,.\label{5dpoinc38}
\end{eqnarray}
Using in addition (\ref{5dpoinc24}), we rewrite (\ref{5dpoinc37}) and (\ref{5dpoinc38}) as
\begin{eqnarray}
\delta\,e_\alpha (x\,,\xi\,'\,) &=& \beta _\alpha{}^\beta\,e_\beta (x\,,\xi\,'\,)\,,\label{5dpoinc39}\\
\delta\,p\,(x\,,\xi\,'\,) &=& 0\,,\label{5dpoinc40}
\end{eqnarray}
where, as before, it is understood that $\beta _\alpha{}^\beta =\beta _\alpha{}^\beta (x)$. The remarkable result (\ref{5dpoinc40}) shows that (\ref{5dpoinc23}) provides a Poincar\'e invariant description of position.

Comparing the position structure (\ref{5dpoinc23}) with the variations (\ref{5dpoinc39}) and (\ref{5dpoinc40}), together with the gauge transformations (\ref{5dpoinc42}) of $\xi ^\mu$, calculated independently from the simplified form (\ref{nlr04}) of the general transformation formula (\ref{secttrans2}) (or (\ref{nonlintrans1})), we find the consistence condition
\begin{equation}
\delta\,\mathfrak{o}=\epsilon ^\mu\,e_\mu \,,\label{5dpoinc43}
\end{equation}
(similar to (\ref{5dpoinc29})) showing that the basis vectors (\ref{5dpoinc16bis}) on $\Sigma $ behave analogously to (\ref{5dpoinc11}) on $M$ under gauge transformations.

\subsection{Lateral displacement of 5-D frames}

As we know, with respect to the underlying base space $M$, gauge transformations are vertical automorphisms moving sections along fibers placed over fixed points $x\in M$. Next we define a lateral link between sections sited on fibres attached to different points of the base space. For that purpose, with the help of connections --defining horizontality--, we introduce parallel transport of fibers as follows \cite{Nakahara}.

In a principal bundle $P(M\,,G)$, let $\gamma (\lambda )$ be a curve in the base space $M$. Horizontal lifts of $\gamma (\lambda )$ are curves $\overline{\gamma}(\lambda )$ in $P$ lying over $\gamma $ in the sense that $\pi _{_{PM}} (\overline{\gamma}) =\gamma $, and such that vectors $\overline{X}$ tangent to $\overline{\gamma}$ at any point are horizontal; that is, such that $\overline{X}\rfloor\omega =0$. For a given curve $\gamma $, there exists only one horizontal lift $\overline{\gamma}$ passing through a given initial point $u_0\in\pi ^{-1}(\gamma (0))$. Parallel transport of a fiber point $u\in\pi ^{-1}(x)$ consists in its mapping to a different point $u\,'\in\pi ^{-1}(x\,' )$, located on a neighboring fiber, {\it moving} along the unique horizontal lift $\overline{\gamma}$ joining both points.

In order to operate with such lifts, we take them to be $\overline{\gamma}(\lambda )\equiv s_{_{MP}}(\lambda ) = R_{\tilde{g}}\circ\sigma _{_{MP}}(\gamma (\lambda ))$, identical with parametrized sections (\ref{locsect1}), where $\tilde{g}=\tilde{g}(\gamma (\lambda ))$ and $\tilde{g}(\gamma (0)) =\tilde{e}\in G$, so that $\overline{\gamma}(0)\equiv s_{_{MP}}(0) = \sigma _{_{MP}}(\gamma (0))$. Then, provided $X$ is a tangent vector to $\gamma (\lambda )$ at $\gamma (0)$, the pushed-forward vector $\overline{X}:= s_{_{MP}*}\,X$ is tangent to $\overline{\gamma}(\lambda )$ at $\overline{\gamma}(0)$. Its horizontality condition reads $s_{_{MP}*}\,X\rfloor\omega =0$.

Let us first apply this result to the standard Poincar\'e principal bundle. It is easy to check that (\ref{poinc02}) with (\ref{poinc03}) and (\ref{poinc03bis}), in the 5-dimensional matrix representation (\ref{5dpoinc07}) (and (\ref{5dpoinc09})), can be expressed as
\begin{eqnarray}
(\Theta _{_G})_A{}^B &=&\,i\,u^{\,\lambda\alpha} du_\lambda {}^\beta\,(\Lambda _{\alpha\beta})_A{}^B -i\,d\xi ^\lambda u_\lambda{}^\mu\,(P_\mu )_A{}^B\nonumber\\
&=& (\tilde{g}^{-1})_A{}^C\,d\,\tilde{g}_C{}^B\,,\label{5dpoinc44}
\end{eqnarray}
so that, in this particular representation, the connection form (\ref{compconnform3}) becomes
\begin{equation}
\omega _A{}^B =\,(\tilde{g}^{-1})_A{}^C\,d\tilde{g}_C{}^B + (\tilde{g}^{-1})_A{}^C\,\pi _{_{PM}}^*\,(A_{_M})_C{}^D\,\tilde{g}_D{}^B\,,\label{5dpoinc45}
\end{equation}
where $(A_{_M})_C{}^D$ is the 5-dimensional representation of (\ref{pbconn04}). From the horizontality condition $s_{_{MP}*}\,X\rfloor \omega _A{}^B =0$ with (\ref{5dpoinc45}) follows the parallel transport equation
\begin{equation}
s_{_{MP}*}\,X\rfloor\,d(\tilde{g}^{-1})_A{}^B = (\tilde{g}^{-1})_A{}^C\,( X\rfloor A_{_M})_C{}^B\,.\label{5dpoinc46}
\end{equation}
Following \cite{Nakahara}, let us now combine the result (\ref{5dpoinc46}) with the following considerations on the vector basis (\ref{5dpoinc11}). Showing the $\gamma (\lambda )$-curve-dependence with the help of a simplified notation where only the curve parameter $\lambda$ is displayed, we start rewriting the equivalence class (\ref{5dpoinc11}) in terms of a convenient representative as
\begin{eqnarray}
e_A(\lambda ) &:=& \left|\left(\,\sigma _{_{MP}}(\lambda )\,,{\buildrel o\over{e}}_A\,\right)\right|\nonumber\\
&=& \left|\left(\,R_{\tilde{g}}\circ\sigma _{_{MP}}(\lambda )\,,(\tilde{g}^{-1})_A{}^B\,{\buildrel o\over{e}}_B\,\right)\right|\nonumber\\
&=& \left|\left(\,s_{_{MP}}(\lambda )\,,(\tilde{g}^{-1})_A{}^B\,{\buildrel o\over{e}}_B\,\right)\right|\,.\label{5dpoinc47}
\end{eqnarray}
Condition (\ref{5dpoinc46}) guiding parallel transport of fibres induces a lateral displacement of frames. Indeed, let us introduce a differential operator $\nabla _X\,$ defined by a suitable differential action on the second element of the ordered-pair of the last expression in (\ref{5dpoinc47}), namely
\begin{equation}
\nabla _X e_A(\lambda )\Big|_{\lambda =0} := \left|\left( s_{_{MP}}(\lambda )\,,-s_{_{MP}*} X\rfloor d (\tilde{g}^{-1})_A{}^B\,{\buildrel o\over{e}}_B\,\right)\right|_{\lambda =0}\,.\label{5dpoinc48}
\end{equation}
Replacing (\ref{5dpoinc46}) in (\ref{5dpoinc48}) and taking into account that $(\tilde{g}^{-1})_A{}^C (\gamma (0)) =\delta _A^C $, we get
\begin{eqnarray}
\nabla _X e_A(\lambda )\Big|_{\lambda =0} &=& \left|\left( s_{_{MP}}(\lambda )\,,-(\tilde{g}^{-1})_A{}^C ( X\rfloor A_{_M})_C{}^B\,{\buildrel o\over{e}}_B\,\right)\right|_{\lambda =0}\nonumber\\
&=& \left|\left(\,\sigma _{_{MP}}(0)\,,-( X\rfloor A_{_M})_A{}^B\,{\buildrel o\over{e}}_B\,\right)\right|\nonumber\\
&=& -( X\rfloor A_{_M})_A{}^B\,e_B (0)\,,\label{5dpoinc49}
\end{eqnarray}
which, in terms of the 5-D representation of the connection (\ref{pbconn04}), reduces to
\begin{eqnarray}
\nabla _X\, e_A(\lambda )\Big|_{\lambda =0} &=& i\,X\rfloor\Bigl(\,\Gamma ^{\alpha\beta}(\Lambda _{\alpha\beta} )_A{}^B\nonumber\\
&&\hskip1.5cm +{\buildrel (T)\over{\Gamma ^\mu}}\,(P_\mu )_A{}^B\Bigr)\,e_B (0)\nonumber\\
&=& o_{A\alpha}\,( X\rfloor\Gamma ^{\alpha\beta})\,e_\beta (0)\nonumber\\
&&\hskip0.2cm + l^{-1}\,\delta _A^5\,( X\rfloor {\buildrel (T)\over{\Gamma ^\mu}})\,e_\mu (0)\,.\label{5dpoinc50}
\end{eqnarray}
Eq.(\ref{5dpoinc50}) decomposed in components and generalized to any point yields
\begin{eqnarray}
\nabla _X\,e_\alpha (x)&=& ( X\rfloor\Gamma _\alpha{}^\beta )\,e_\beta (x)\,,\label{5dpoinc51}\\
\nabla _X\,e_5 (x)&=& l^{-1}\,( X\rfloor {\buildrel (T)\over{\Gamma ^\mu}})\,e_\mu (x)\,.\label{5dpoinc52}
\end{eqnarray}
By defining the $X$-independent nabla operator $\nabla $ by
\begin{equation}
X\rfloor\nabla e_A(x) =\nabla _X\, e_A(x)\,,\label{5dpoinc53}
\end{equation}
from (\ref{5dpoinc51}) and (\ref{5dpoinc52}) we get
\begin{eqnarray}
\nabla e_\alpha (x)&=& \Gamma _\alpha{}^\beta \,e_\beta (x)\,,\label{5dpoinc54}\\
\nabla e_5 (x)&=& l^{-1}\,{\buildrel (T)\over{\Gamma ^\mu}}\,e_\mu (x)\,.\label{5dpoinc55}
\end{eqnarray}

Something similar can be done --to some extent-- for the basis vectors (\ref{5dpoinc16}) on the intermediate base space $\Sigma$. In this case we make use of the connection 1-form (\ref{connform1}), which in the 5-dimensional representation reads
\begin{equation}
\omega _A{}^B =\,( a^{-1})_A{}^C\,d a_C{}^B + ( a^{-1})_A{}^C\,\pi _{_{P\Sigma}}^*\,(\Gamma _{_\Sigma})_C{}^D\,a_D{}^B\,.\label{5dpoinc56}
\end{equation}
The parallel transport equation analogous to (\ref{5dpoinc46}) is found with the help of a vector $Y$ on $\Sigma$ which is required to be horizontal, that is, such that it satisfies $s_{_{\Sigma P}*}\,Y\rfloor \omega _A{}^B =0\,$, yielding
\begin{equation}
s_{_{\Sigma P}*}\,Y\rfloor\,d( a^{-1})_A{}^B = ( a^{-1})_A{}^C\,( Y\rfloor\Gamma _{_\Sigma})_C{}^B\,.\label{5dpoinc57}
\end{equation}
Following similar steps as before in (\ref{5dpoinc47}), we transform (\ref{5dpoinc16}) as
\begin{eqnarray}
\widehat{e}_A(x\,,\xi\,) &:=& \left|\left(\,\sigma _{_{\Sigma P}}( x\,,\xi )\,,{\buildrel o\over{\widehat{e}}}_{\,A}\,\right)\right|\nonumber\\
&=& \left|\left(\,R_a\circ\sigma _{_{\Sigma P}}( x\,,\xi )\,,( a^{-1})_A{}^B\,{\buildrel o\over{\widehat{e}}}_{\,B}\,\right)\right|\nonumber\\
&=& \left|\left(\,s_{_{\Sigma P}}( x\,,\xi )\,,( a^{-1})_A{}^B\,{\buildrel o\over{\widehat{e}}}_{\,B}\,\right)\right|\,,\label{5dpoinc58}
\end{eqnarray}
and we introduce the (\ref{5dpoinc48})-analogous definition of the nabla operator
\begin{equation}
\nabla _Y\widehat{e}_A(\lambda )\Big|_{\lambda =0} := \left|\left( s_{_{\Sigma P}}(\lambda )\,,-s_{_{\Sigma P}*}Y\rfloor d ( a^{-1})_A{}^B\,{\buildrel o\over{\widehat{e}}}_B\right)\right|_{\lambda =0}\,,\label{5dpoinc59}
\end{equation}
which together with condition (\ref{5dpoinc57}), proceeding as in (\ref{5dpoinc49}), yields
\begin{equation}
\nabla _Y\,\widehat{e}_A(\lambda )\Big|_{\lambda =0} = -( Y\rfloor\Gamma _{_\Sigma})_A{}^B\,\widehat{e}_B (0)\,.\label{5dpoinc60}
\end{equation}
Replacing in (\ref{5dpoinc60}) the 5-D representation of connection (\ref{poincconnform3}), we get
\begin{eqnarray}
\nabla _Y\,\widehat{e}_A(\lambda )\Big|_{\lambda =0} &=& o_{A\alpha}\,( Y\rfloor\pi _{_{\Sigma M}}^*\,\Gamma ^{\alpha\beta})\,\widehat{e}_\beta (0)\nonumber\\
&&\hskip0.2cm  + l^{-1}\,\delta _A^5\,( Y\rfloor\vartheta _{_\Sigma}^\mu\,)\,\widehat{e}_\mu (0)\,,\label{5dpoinc61}
\end{eqnarray}
which, in components and generalized to any point, reads
\begin{eqnarray}
\nabla _Y\,\widehat{e}_\alpha (x\,,\xi\,) &=& ( Y\rfloor\pi _{_{\Sigma M}}^*\,\Gamma _\alpha{}^\beta )\,\widehat{e}_\beta (x\,,\xi\,)\,,\label{5dpoinc62}\\
\nabla _Y\,\widehat{e}_5 (x\,,\xi\,) &=& l^{-1}\,( Y\rfloor \vartheta _{_\Sigma}^\mu\,)\,\widehat{e}_\mu (x\,,\xi\,)\,.\label{5dpoinc63}
\end{eqnarray}
Here we have to notice that, strictly speaking, (\ref{5dpoinc62}) is the only rigorous result we can derive from (\ref{5dpoinc59}), due to the fact that both members of (\ref{5dpoinc63}) vanish for $Y$ horizontal. (This follows from the structure (\ref{5dpoinc05}) of $a_A{}^B$ in (\ref{5dpoinc56}) --where $d a_5{}^\mu =0\,$-- and from (\ref{variety08}) below, respectively.) Notwithstanding, we keep the formal expression (\ref{5dpoinc63}) as a guide. In terms of the $Y$-independent operator $\nabla $ such that
\begin{equation}
Y\rfloor\nabla\widehat{e}_A (x\,,\xi\,) =\nabla _Y\,\widehat{e}_A (x\,,\xi\,)\,,\label{5dpoinc64}
\end{equation}
from (\ref{5dpoinc62}) we get
\begin{equation}
\nabla e_\alpha (x\,,\xi\,) = \pi _{_{\Sigma M}}^*\,\Gamma _\alpha{}^\beta \,e_\beta (x\,,\xi\,)\,,\label{5dpoinc65}
\end{equation}
whereas from (\ref{5dpoinc63}) we cannot conclude anything. However, in order to complete the action of the nude nabla operator on the components of (\ref{5dpoinc24}), we postulate --consistently with (\ref{5dpoinc63})-- that
\begin{equation}
\nabla p\,(x\,,\xi\,) = \vartheta _{_\Sigma}^\mu\,e_\mu (x\,,\xi\,)\,,\label{5dpoinc66}
\end{equation}
where definition (\ref{5dpoinc22}) is used. (See Refs. \cite{Gronwald:1997bx} \cite{Cartan}.) As an argument in support of  (\ref{5dpoinc66}), let us compare it with the result of acting with the operator $\nabla $ on (\ref{5dpoinc23}) taking (\ref{5dpoinc65}) into account, that is
\begin{eqnarray}
\nabla p &=& \nabla (\,\mathfrak{o} +\xi ^\mu\,e_\mu\,)\nonumber\\
&=& \nabla\mathfrak{o} + d\xi ^\mu\,e_\mu + \xi ^\mu\,\nabla e_\mu\nonumber\\
&=& \nabla\mathfrak{o} + (\,d\xi ^\mu +\pi _{_{\Sigma M}}^*\,\Gamma _\nu{}^\mu\,\xi ^\nu\,)\,e_\mu\,.\label{5dpoinc67}
\end{eqnarray}
Checking the r.h.s. of (\ref{5dpoinc67}) against definition (\ref{connform4}) of $\vartheta _{_\Sigma}^\mu$, we realize that the only remaining requirement for the validity of (\ref{5dpoinc66}) is the condition
\begin{equation}
\nabla \mathfrak{o} = \pi _{_{\Sigma M}}^*\,{\buildrel (T)\over{\Gamma ^\mu}}\,e_\mu\,,\label{5dpoinc68}
\end{equation}
similar to the already deduced result (\ref{5dpoinc55}).

\section{Tangent space $T(P)$ of the Poincar\'e composite bundle }

Returning to the bundle manifold, we next organize the tangent bundle identifying horizontal and vertical vectors corresponding to each of the two bundle sectors (\ref{compbundle02}) and (\ref{compbundle03}). With the help of connection (\ref{variety18})--(\ref{variety20}) introduced in Section IV, we start looking for horizontal and vertical vectors in the tangent space $T(\Sigma )$ of the {\it intermediate space} $\Sigma $.

\subsection{Vectors of the intermediate tangent space $T(\Sigma )$}

Let us construct a horizontal vector on $T(\Sigma )$ (using only the orbital part (\ref{poinc14}) of the angular momentum generators $\overline{L}^{_{(\Lambda )}}_{\alpha\beta }$ since the intrinsic part is not defined on $\Sigma$). We write the horizontal vector in different forms as
\begin{eqnarray}
{\buildrel {_\Sigma}\over{X}}_i &:=&\sigma _{{_{M\Sigma}}*}\,\partial _i +i\,\pi _{_{P\Sigma}*}\,( {\buildrel (T)\over{\Gamma _i\,^\mu}}\overline{L}^{_{(P)}}_\mu +\Gamma _i^{\alpha\beta}\overline{L}^{_{(Orb)}}_{\alpha\beta }\,)\nonumber\\
&=&\sigma _{{_{M\Sigma}}*}\,\partial _i -( \Gamma _{i\nu}{}^\mu\xi ^\nu +{\buildrel (T)\over{\Gamma _i\,^\mu}}\,)\,\pi _{_{P\Sigma}*}\,\partial _{\xi ^\mu}\,,\nonumber\\
&=& s_{{_{M\Sigma}}*}\,\partial _i - e_i{}^\mu\,\pi _{_{P\Sigma}*}\,\partial _{\xi ^\mu}\,.\label{variety07}
\end{eqnarray}
where we used the relation $\,s_{{_{M\Sigma}}*}\,\partial _i = \sigma _{{_{M\Sigma}}*}\,\partial _i +\partial _i\xi ^\mu\,\pi _{_{P\Sigma}*}\,\partial _{\xi ^\mu}\,$ and definition (\ref{variety06}). Vector (\ref{variety07}) contracted with (\ref{connform4}) yields
\begin{equation}
{\buildrel {_\Sigma}\over{X}}_i \rfloor \vartheta _{_{\Sigma}}^\mu =0\,,\label{variety08}
\end{equation}
implying that also its contraction with (\ref{variety19}) vanishes, that is
\begin{equation}
\sigma _{{_{\Sigma P}}*}\,{\buildrel {_\Sigma}\over{X}}_i\,\rfloor {\buildrel (P)\over{\omega }} =0\,,\label{variety21}
\end{equation}
so that, in fact, ${\buildrel {_\Sigma}\over{X}}_i$ is the horizontal vector corresponding to the translational connection part (\ref{variety19}).

On the other hand, we recognize in (\ref{poinc06}) the fundamental vector pointing vertically along the fibers $G/H$ of the fibred space $\Sigma\rightarrow M$. To check it, we rewrite (\ref{variety19}) using (\ref{auxil01}) and defining
\begin{equation}
\vartheta _{_P}^\mu :=\pi _{_{P\Sigma}}^* \vartheta _{_{\Sigma}}^\mu = d_{_P}\xi ^\mu + \pi _{_{PM}}^* dx^i (\Gamma _{i\nu}{}^\mu \xi\,^\nu +{\buildrel (T)\over{\Gamma _i\,^\mu}}\,)\,,\label{connform4bis}
\end{equation}
(compare with (\ref{connform4})), and establishing the chain of equalities
\begin{equation}
\delta _\mu ^\nu = \partial _{\xi ^\mu}\rfloor d_{_P}\xi ^\nu = \partial _{\xi ^\mu}\rfloor \pi _{_{P\Sigma}}^*\,d_{_\Sigma}\xi ^\nu\,,\label{auxil02}
\end{equation}
so that, contracting (\ref{poinc06}) with (\ref{variety19}), we get
\begin{equation}
L^{_{(P)}}_\mu\rfloor {\buildrel (P)\over {\omega }} = P_\mu\,.\label{variety22}
\end{equation}
However, for later convenience, we replace the fundamental vector with a modified one. Let us define the latter as
\begin{equation}
{\buildrel {_\Sigma}\over{e}}_\mu := e_\mu{}^i\,s_{{_{M\Sigma}}*}\,\partial _i\,.\label{variety14}
\end{equation}
Despite its lack of apparent relationship to the standard fundamental vector $L^{_{(P)}}_\mu\sim\partial _{\xi ^\mu}$, introducing the inverse matrix $e_\mu{}^i$ of (\ref{variety06}), such that $e_\mu{}^i\,e_i{}^\nu =\delta _\mu ^\nu $, and multiplying by it (\ref{variety07}), we find (\ref{variety14}) to be equal to
\begin{equation}
{\buildrel {_\Sigma}\over{e}}_\mu = \pi _{_{P\Sigma}*}\,\partial _{\xi ^\mu} +e_\mu{}^i {\buildrel {_\Sigma}\over{X}}_i\,,\label{variety10}
\end{equation}
mainly differing from the fundamental vector (\ref{poinc06})$\sim\partial _{\xi ^\mu}$ by a term proportional to the horizontal vector (\ref{variety07}) satisfying (\ref{variety21}). In analogy to the standard fundamental vector (\ref{poinc06}), we realize that a vector constructed from (\ref{variety10}) satisfies a condition similar to (\ref{variety22}), namely
\begin{equation}
i\,\sigma _{{_{\Sigma P}}*}\,{\buildrel {_\Sigma}\over{e}}_\mu \rfloor {\buildrel (P)\over {\omega }} = P_\mu\,,\label{variety23}
\end{equation}
or, in a more intuitive formulation, for (\ref{variety10}) itself holds
\begin{equation}
{\buildrel {_\Sigma}\over{e}}_\mu\rfloor \vartheta _{_{\Sigma}}^\nu =\delta _\mu ^\nu\,.\label{variety11}
\end{equation}
(Compare with (\ref{variety08}).) In analogy to (\ref{variety05}), with the help of the inverse matrix $e_\mu{}^i$ of (\ref{variety06}) we introduce in the base space $M$ the basis vectors
\begin{equation}
{\buildrel {_M}\over{e}}_\mu := e_\mu{}^i\partial _i\,,\label{variety13}
\end{equation}
which in view of definition (\ref{variety14}) are found to relate to the tangent vectors (\ref{variety10}) of $T(\Sigma )$ as
\begin{equation}
{\buildrel {_\Sigma}\over{e}}_\mu = s_{{_{M\Sigma}}*}\,{\buildrel {_M}\over{e}}_\mu\,.\label{variety14bis}
\end{equation}
That is, ${\buildrel {_\Sigma}\over{e}}_\mu $ is the push-forward of ${\buildrel {_M}\over{e}}_\mu $ by $s_{{_{M\Sigma}}}$. Observe that, recalling (\ref{variety05}), (\ref{variety11}) and (\ref{variety14bis}), basis vectors (\ref{variety13}) and 1-forms (\ref{variety05}), as much as their related quantities (\ref{variety10}) and (\ref{connform4}), satisfy
\begin{eqnarray}
{\buildrel {_M}\over{e}}_\mu\rfloor \vartheta _{_M}^\nu &=& {\buildrel {_M}\over{e}}_\mu\rfloor s_{_{M\Sigma}}^*\vartheta _{_{\Sigma}}^\nu \nonumber\\
&=& s_{{_{M\Sigma}}*}\,{\buildrel {_M}\over{e}}_\mu\rfloor \vartheta _{_{\Sigma}}^\nu \nonumber\\
&=& {\buildrel {_\Sigma}\over{e}}_\mu\rfloor \vartheta _{_{\Sigma}}^\nu = \delta _\mu ^\nu\,,\label{variety15}
\end{eqnarray}
(see Appendix D), so that, since we identified the 1-forms $\vartheta _{_M}^\mu$ in (\ref{variety05}) as ordinary tetrads, from (\ref{variety15}) we find that $\{\,{\buildrel {_M}\over{e}}_\mu\,\}$ is the ordinary {\it vierbein} vector basis on $M$.

\subsection{Other vectors of the tangent bundle $T(P)$}

For the sake of completeness, we introduce the vectors corresponding to the bundle sector $P\rightarrow\Sigma$ although they do not play any role in what follows. In the mentioned principal subbundle (\ref{compbundle02}), besides the vertical (fundamental) vector $L^{_{(\Lambda )}}_{\alpha\beta }$ such that
\begin{equation}
L^{_{(\Lambda )}}_{\alpha\beta }\,\rfloor{\buildrel (\Lambda )\over{\omega}} =\Lambda _{\alpha\beta }\,,\label{variety24}
\end{equation}
one can define a horizontal vector built in terms of (\ref{variety10}) as
\begin{equation}
{\buildrel {_P}\over{E}}_\mu := i\,\left(\,\sigma _{{_{\Sigma P}}*}{\buildrel {_\Sigma}\over{e}}_\mu +i\,\Gamma _\mu ^{\alpha\beta}\,\overline{L}^{_{(Int)}}_{\alpha\beta }\,\right)\,,\label{variety26}
\end{equation}
(with the intrinsic part of $\overline{L}^{_{(\Lambda )}}_{\alpha\beta }$ as defined in (\ref{poinc12}), (\ref{poinc14}) and with the Lorentz connection components $\Gamma _\mu^{\alpha\beta}:= e_\mu{}^i\,\Gamma _i^{\alpha\beta}\,$) satisfying, when contracted with (\ref{variety19}) and (\ref{variety20}) respectively
\begin{eqnarray}
{\buildrel {_P}\over{E}}_\mu\rfloor {\buildrel (P)\over {\omega }} &=& P_\mu\,,\label{variety28}\\
{\buildrel {_P}\over{E}}_\mu\rfloor {\buildrel (\Lambda )\over{\omega}} &=& 0\,,\label{variety28bis}
\end{eqnarray}
being (\ref{variety28bis}) the horizontality condition for the subbundle  $P\rightarrow\Sigma$. To check (\ref{variety28bis}), one must take into account that $\overline{L}^{_{(Int)}}_{\alpha\beta}\rfloor\overline{\Theta}_{^{(\Lambda )}}^{\mu\nu} =\delta _{[\alpha}^\mu \delta _{\beta ]}^\nu $ and
\begin{eqnarray}
\sigma _{{_{\Sigma P}}*}\,{\buildrel {_\Sigma}\over{e}}_\mu\rfloor\pi _{_{PM}}^*\Gamma ^{\alpha\beta} &=&
{\buildrel {_\Sigma}\over{e}}_\mu\rfloor\sigma _{{_{\Sigma P}}}^*\,\pi _{_{P\Sigma}}^*\,\pi _{_{\Sigma M}}^*\Gamma ^{\alpha\beta}\nonumber\\
&=& s_{{_{M\Sigma}}*}\,{\buildrel {_M}\over{e}}_\mu\rfloor\pi _{_{\Sigma M}}^*\Gamma ^{\alpha\beta}\nonumber\\
&=& {\buildrel {_M}\over{e}}_\mu\rfloor\Gamma ^{\alpha\beta} = e_\mu{}^i\,\Gamma _i^{\alpha\beta}\nonumber\\
&=:& \Gamma _\mu^{\alpha\beta}\,.\label{variety29}
\end{eqnarray}
A scheme of the basic vectors of the tangent bundle $T(P)$ is displayed in Fig.3.
\begin{figure}[h]
\begin{center}
\includegraphics[totalheight=63mm,width=89mm]{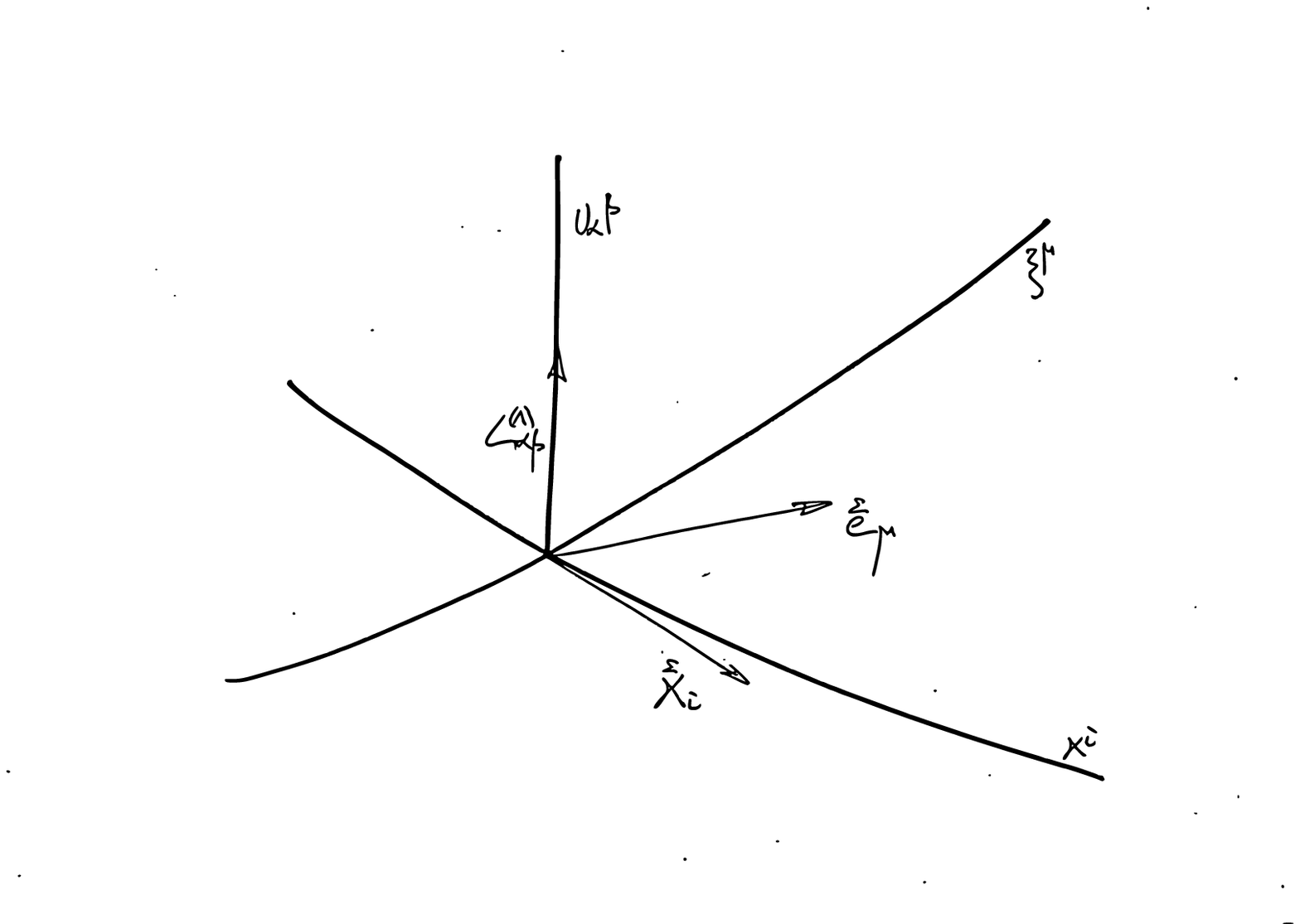}
\caption{\label{Fig3}Tangent bundle $T(P)$}
\end{center}
\end{figure}

The composite fiber bundle structure is compatible with the existence of horizontal and vertical vectors corresponding to the ordinary principal fiber bundle $P(M\,,G)$. Indeed, the vector built in analogy to (\ref{variety07})
\begin{equation}
{\buildrel {_P}\over{X}}_i := \sigma _{{_{MP}}*}\,\partial _i +i\,\,( {\buildrel (T)\over{\Gamma _i\,^\mu}}\overline{L}^{_{(P)}}_\mu +\Gamma _i^{\alpha\beta}\overline{L}^{_{(\Lambda )}}_{\alpha\beta }\,)
 \,,\label{variety30}
\end{equation}
is such that
\begin{equation}
{\buildrel {_P}\over{X}}_i\rfloor {\buildrel (P)\over{\omega}} = 0\,,\qquad {\buildrel {_P}\over{X}}_i\rfloor {\buildrel (\Lambda )\over{\omega}} = 0\,,\label{variety31}
\end{equation}
so that it is horizontal in the total bundle; the corresponding (vertical) fundamental vector is a combination of (\ref{variety22}) and (\ref{variety24}).

\section{Considerations on the tangent space $T(\Sigma )$}

\subsection{Enlarging the frames of the tangent bundle with an origin}

The natural basis of the tangent bundle $T(\Sigma )$ is $\{\,\sigma _{{_{M\Sigma}}*}\,\partial _i\,,\pi _{_{P\Sigma}*}\,\partial _{\xi ^\mu}\,\}$. In Section VI A, with the help of the translational connection contribution ${\buildrel (P)\over\omega}$ (involving the tetrads $\vartheta _{_\Sigma}^\mu$ on $\Sigma$), we first divided the tangent space $T(\Sigma )$ into a horizontal subspace with basis $\{\,{\buildrel {_\Sigma}\over{X}}_i\,\}$ and a vertical subspace with basis $\{\,\pi _{_{P\Sigma}*}\,\partial _{\xi ^\mu}\,\}$, and then we replaced the latter vector basis by the modified one $\{\,{\buildrel {_\Sigma}\over{e}}_\mu\,\}$, relating as (\ref{variety14bis}) to ordinary {\it vierbeins}, which we proved in \cite{Tresguerres:2002uh} to transform as Lorentz vectors. This suggests to identify $\{\,{\buildrel {_\Sigma}\over{e}}_\mu\,\}$ as a Lorentz frame, and the set of all such frames at all points of $\Sigma$ as a frame bundle over $\Sigma$ with the Lorentz group as its structure group.

Next, following \cite{Kobayashi:1963}, see pgs. 125 ff., {\it we enlarge the bundle of linear (Lorentz) frames to the bundle of affine (Poincar\'e) frames} adding a point ${\buildrel {_\Sigma}\over{\mathfrak{o}}}\in T(\Sigma )$ to the Lorentz frame $\{\,{\buildrel {_\Sigma}\over{e}}_\mu\,\}$ to complete a Poincar\'e frame. By later convenience, we denote this point like in (\ref{5dpoinc21}) as $l^{-1}{\buildrel {_\Sigma}\over{\mathfrak{o}}}$, so that the Poincar\'e frames, which constitute a principal bundle with $G=$Poincar\'e as its structure group \cite{Kobayashi:1963}, take the form
\begin{equation}
{\buildrel {_\Sigma}\over{e}}_A = \binom{{\buildrel {_\Sigma}\over{e}}_\alpha }{l^{-1}{\buildrel {_\Sigma}\over{\mathfrak{o}}}\,}\,,\label{5dpoinc72a}
\end{equation}
similar to the basis (\ref{5dpoinc16bis}) built from the natural basis $\{\,{\buildrel o\over{e}}_{\,A}\}$ of $\mathbb{R}^5$.

In analogy to (\ref{5dpoinc18}), where we recognized a relation of the type (\ref{nlr06}) characteristic for representation fields of NLR's, from (\ref{5dpoinc72a}) we also define
\begin{equation}
{\buildrel {_\Sigma}\over{\widehat{e}}}_A = (b^{-1})_A{}^B\,{\buildrel {_\Sigma}\over{e}}_B = \binom{{\buildrel {_\Sigma}\over{e}}_\alpha }{l^{-1}\,{\buildrel {_\Sigma}\over{p}}\,}\,,\label{5dpoinc72b}
\end{equation}
where ${\buildrel {_\Sigma}\over{p}}$ is given by (\ref{5dpoinc23}). The Poincar\'e frame (\ref{5dpoinc72b}) closely resembles the 5-dimensional vector basis (\ref{5dpoinc24}), both being defined on the $\Sigma $-manifold.

As revealed by the 5-D associated vector bundle, although the fifth basis component is an elusive quantity which seems to be hidden in the tangent space $T(\Sigma )$ (maybe because no vector proportional to $\partial _{\xi ^5}$ exists), it is reasonable to consider it as an unavoidable constitutive piece of a complete vector representation space of the Poincar\'e group.

Even if not explicitly present in the tangent bundle formalism, the presence of $e_5$ can be recognized by the traces it leaves behind on other mathematical objects, in particular through translational connections. According to (\ref{5dpoinc66}), its mark is present in the internal structure (\ref{connform4}) of the modified translational connection $\vartheta _{_\Sigma}^\mu $. A fundamental correspondence seems to exist between the associated 5-D vector space and the part of the tangent space of $T(\Sigma )$ expanded by (\ref{5dpoinc72b}).

\subsection{Conjecture}

Despite the incompleteness of the scheme, mainly due to the difficulty in detecting the presence of ${\buildrel {_\Sigma}\over{\mathfrak{o}}}$ or ${\buildrel {_\Sigma}\over{p}}$ in $T(\Sigma )$, we propose to identify (\ref{5dpoinc72a}) with the vector basis (\ref{5dpoinc16bis}) of the 5-dimensional representation space of the Poincar\'e group, and its (\ref{5dpoinc18})-analogous modification (\ref{5dpoinc72b}) with (\ref{5dpoinc24}). In other words, we postulate that (\ref{5dpoinc72a}) is a basis of a 5-D vector space constituting a linear representation space of the 5-D matrix representation of the Poincar\'e group, being (\ref{5dpoinc72b}) its (\ref{nlr06})-type modification associated to NLR's.

Accordingly, in the following we do not distinguish (\ref{5dpoinc72b}) from (\ref{5dpoinc24}), so that we can apply to the tangent space quantities the features previously found for (\ref{5dpoinc24}). So, we assume the Poincar\'e frame (\ref{5dpoinc72b}) to consist of the basis vectors ${\buildrel {_\Sigma}\over{e}}_\alpha $ and the position point
\begin{equation}
{\buildrel {_\Sigma}\over{p}} = {\buildrel {_\Sigma}\over{\mathfrak{o}}} +\xi ^\mu\,{\buildrel {_\Sigma}\over{e}}_\mu\,,\label{5dpoinc73}
\end{equation}
built in analogy to (\ref{5dpoinc23}). The gauge transformation of the different pieces of (\ref{5dpoinc73}) under the Poincar\'e group, according to (\ref{5dpoinc39}), (\ref{5dpoinc40}), (\ref{5dpoinc43}) and (\ref{5dpoinc42}), are respectively
\begin{eqnarray}
\delta\,{\buildrel {_\Sigma}\over{e}}_\mu &=& \beta _\mu{}^\nu\,{\buildrel {_\Sigma}\over{e}}_\nu\,,\label{5dpoinc74}\\
\delta\,{\buildrel {_\Sigma}\over{\mathfrak{o}}} &=& \epsilon ^\mu\,{\buildrel {_\Sigma}\over{e}}_\mu \,,\label{5dpoinc75}\\
\delta\xi ^\mu &=&-\xi ^\nu\beta _\nu{}^\mu -\epsilon ^\mu\,,\label{5dpoinc76}\\
\delta\,{\buildrel {_\Sigma}\over{p}} &=& 0\,.\label{5dpoinc77}
\end{eqnarray}
On the other hand, their lateral displacements, according to (\ref{5dpoinc65}), (\ref{5dpoinc66}) and (\ref{5dpoinc68}), read
\begin{eqnarray}
\nabla\,{\buildrel {_\Sigma}\over{e}}_\mu &=& \pi _{_{\Sigma M}}^*\,\Gamma _\mu{}^\nu\,{\buildrel {_\Sigma}\over{e}}_\nu\,,\label{5dpoinc78}\\
\nabla\,{\buildrel {_\Sigma}\over{\mathfrak{o}}} &=& \pi _{_{\Sigma M}}^*\,{\buildrel (T)\over{\Gamma ^\mu}}\,{\buildrel {_\Sigma}\over{e}}_\mu \,,\label{5dpoinc79}\\
\nabla\,{\buildrel {_\Sigma}\over{p}} &=& \nabla (\,{\buildrel {_\Sigma}\over{\mathfrak{o}}} +\xi ^\mu\,{\buildrel {_\Sigma}\over{e}}_\mu\,)\nonumber\\
&=& \nabla {\buildrel {_\Sigma}\over{\mathfrak{o}}} + d_{_\Sigma}\xi ^\mu\,{\buildrel {_\Sigma}\over{e}}_\mu + \xi ^\mu\,\nabla {\buildrel {_\Sigma}\over{e}}_\mu\nonumber\\
&=& \bigl[\,d_{_\Sigma}\xi ^\mu +\pi _{_{\Sigma M}}^*\,(\,\Gamma _\nu{}^\mu\,\xi ^\nu +{\buildrel (T)\over{\Gamma ^\mu}}\,)\,\bigr]{\buildrel {_\Sigma}\over{e}}_\mu\nonumber\\
&=:& \vartheta _{_\Sigma}^\mu\,{\buildrel {_\Sigma}\over{e}}_\mu \,.\label{5dpoinc80}
\end{eqnarray}
In addition, in view of (\ref{variety15})
\begin{equation}
{\buildrel {_\Sigma}\over{e}}_\mu\rfloor \vartheta _{_\Sigma}^\nu = {\buildrel {_M}\over{e}}_\mu\rfloor \vartheta _{_M}^\nu =\delta _\mu ^\nu\,.\label{5dpoinc81}
\end{equation}
In what follows, the position object ${\buildrel {_\Sigma}\over{p}}$ plays a central role in the interpretation of motion. First of all due to its particular structure (\ref{5dpoinc73}), which enables the translational parameters $\xi ^\mu$ to acquire their natural interpretation as position vector components. Moreover, it also explains the form (\ref{connform4}) of $\vartheta _{_\Sigma}^\mu\,$ as derived in (\ref{5dpoinc80}), thus determining the configuration (\ref{fourvelocity}) of fourvelocity components. It also gives a definite meaning to equations (\ref{vel01bis}) and (\ref{eqsmot}) to be studied later.

\subsection{Vertical vs. lateral motions}

Concerning eqs.(\ref{5dpoinc74})--(\ref{5dpoinc80}), let us recall that vertical and lateral motions taking place in the framework of a bundle are essentially different in nature, and this statement holds also for motions {\it vertical} or {\it lateral} relatively to the fibred {\it intermediate space} $\Sigma$.

Vertical motions (\ref{5dpoinc74})--(\ref{5dpoinc76}) of frames (gauge transformations along fibres $G/H$) are kinematical in the sense that they do not involve connections (say forces), but only non-dynamical gauge group parameters such as $\epsilon ^\mu$ and $\beta ^{\alpha\beta}$ reflecting symmetries. They act on reference frames (material or merely ideal) by translating, rotating or boosting them with respect to each other, giving rise to physically equivalent descriptions. This holds for instance for the alternative perspectives, related by (\ref{5dpoinc74})--(\ref{5dpoinc77}), on the position (\ref{5dpoinc73}), which remains invariant while {\it observed} from different reference frames.

On the other hand, dynamical changes (\ref{5dpoinc78}), (\ref{5dpoinc79}) of frames are displacements moving from fiber to fiber by means of the operator $\nabla$, whose action on the frame $({\buildrel {_\Sigma}\over e}_\mu \,,l^{-1}\mathfrak{o})$ brings connections to light. While vertical motions merely modify the point of view, leaving the geometrical object ${\buildrel {_\Sigma}\over p}$ unaffected according to (\ref{5dpoinc77}), lateral motions induce real changes (\ref{5dpoinc80}) on ${\buildrel {_\Sigma}\over p}\,$. With these displacements (\ref{5dpoinc80}), pulled back to $M$ as in (\ref{variety05}), one can build the line element $ds^2 = o_{\alpha\beta}\,\vartheta _{_M}^\alpha\,\vartheta _{_M}^\beta$, and moreover, in the related form (\ref{vel01}) below, it induces time evolution, as we will see immediately.

\section{Time evolution}

The starting point of Einstein's Special Relativity is the recognition of the local character of time. Local clocks are able to measure local time intervals either of static processes or of motions over closed paths, such as the round-trip time of a light signal. But the measurement of time intervals whose initial and final instants occur at separate points of space is in principle operationally meaningless unless one postulates a synchronization (or simultaneity) criterion for distant events. Accordingly, also the speeds of particles or of signals moving from one point to another are in principle empirically not well defined.

In Special Relativity, one cuts the Gordian knot defining the global time of an extended inertial reference frame by means of Einstein's synchronization convention of constant one-way speed of light $c$. Moreover, from the requirement of invariance of $c$ follows that the measurements performed in different inertial frames relate to each other through global Poincar\'e transformations, so that the time of a relatively moving observer, which cannot be directly measured, becomes calculable.

\subsection{Flow of events and its generating vector field}

In General Relativity as much as in its gauge extensions, that is, in curved spacetimes, Einstein's definition of global time is not applicable since inertial frames exist just locally (associated for instance to free falling lifts). Any shared time one may introduce to synchronize not coincident although nearby events can only make sense in limited regions. We take from the literature the idea of an {\it extended reference frame} in the presence of gravity, spanned (locally) over a congruence of timelike curves \cite{Minguzzi:2001ts}--\cite{Minguzzi:2005ag}. A congruence is a family of trajectories --paths or {\it worldlines}-- of a set of neighboring points moving jointly. The tangent vectors to the curves constitute a nowhere vanishing fourvelocity vector field whose integral curves are precisely the worldlines. The congruence of curves determines a flow, consisting in a transformation of the space into itself along the worldlines, being the flow generated by the vector field. Thus, the congruence of curves, as much as the tangent vector field and the flow of events through spacetime, reciprocally imply each other \cite{Crampin:1986}.

To be more precise, we introduce a vector field
\begin{equation}
{\buildrel {_\Sigma}\over{u}} = u^\mu\,{\buildrel {_\Sigma}\over{e}}_\mu\,,\label{time02}
\end{equation}
associated to a set of neighboring observers on $T(\Sigma )$, and we restrict our study to the case of timelike vectors (\ref{time02}). In view of (\ref{variety14bis}), the field (\ref{time02}) is trivially related to the vector
\begin{equation}
{\buildrel {_M}\over{u}} = u^\mu\,{\buildrel {_M}\over{e}}_\mu\,,\label{time02bis}
\end{equation}
(defined on $T(M)$ in terms of ordinary {\it vierbeins}) as
\begin{equation}
{\buildrel {_\Sigma}\over{u}} = s_{{_{M\Sigma}}*}\,{\buildrel {_M}\over{u}}\,.\label{uvectors}
\end{equation}
In the following, we distinguish (\ref{time02}) from (\ref{time02bis}) only when necessary. Otherwise we merely write $u$.

The nowhere vanishing timelike vector field ${\buildrel {_\Sigma}\over{u}}$ generates the flow of events which constitutes the observers' experience of change. Actually, it provides a formalization of both, motion and pure passage of time. To see this, let us express (\ref{time02}) referred to different frames boosted with respect to each other --or redefined using the gauge-like boost transformations (\ref{hatfourvel0}), (\ref{hatfourvela}) below-- as
\begin{equation}
{\buildrel {_\Sigma}\over{u}} = u^\mu\,{\buildrel {_\Sigma}\over{e}}_\mu = \hat{u}^\mu\,{\buildrel {_\Sigma}\over{\hat{e}}}_\mu\,.\label{time02bbis}
\end{equation}
By taking the hat components of the timelike vector to be $\hat{u}^a =0$, $\hat{u}^0\neq 0$, we can write (\ref{time02bbis}) alternatively as
\begin{equation}
{\buildrel {_\Sigma}\over{u}} = \begin{cases} u^\mu\,{\buildrel {_\Sigma}\over{e}}_\mu \\
\hat{u}^{_0} {\buildrel {_\Sigma}\over{\hat{e}}}_{_0}\end{cases}\,,\label{time03}
\end{equation}
where ${\buildrel {_\Sigma}\over{u}} = u^\mu\,{\buildrel {_\Sigma}\over{e}}_\mu $ represents the fourvelocity of an object with respect to the observer while ${\buildrel {_\Sigma}\over{u}} = \hat{u}^{_0} {\buildrel {_\Sigma}\over{\hat{e}}}_{_0}$ reflects the passage of (proper) time --pointing in the direction of the proper time vector ${\buildrel {_\Sigma}\over{\hat{e}}}_{_0}$-- as measured in the frame where the object is {\it at rest}.

Finally, making use once more of (\ref{hatfourvel0}), (\ref{hatfourvela}) below, we find the invariant expression
\begin{equation}
u_\mu u^\mu = \hat{u}_\mu \hat{u}^\mu \,.\label{time04prev}
\end{equation}
For timelike vector fields (\ref{time02}), we can take $\hat{u}^a =0$, so that (\ref{time04prev}) reduces to
\begin{equation}
u_\mu u^\mu = -(\hat{u}^0)^2\,,\label{time04}
\end{equation}
where we recognize the timelikeness condition, with the weight $\hat{u}^0$ of the local rhythm of time (not necessarily a constant) in the role of the velocity of light.

\subsection{Frobenius foliation condition}

Instead of Einstein's global synchronization prescription, we introduce a related local simultaneity convention valid for neighboring observers moving along the flow of events generated by the timelike vector field (\ref{time02}).

With the help of the Frobenius theorem, we perform a local spacetime foliation into simultaneity slices cutting the worldlines orthogonally. In order to carry out the foliation, first one has to introduce a suitable 1-form $\omega _\tau$ satisfying, with the vector field (\ref{time02}), the relation
\begin{equation}
{\buildrel {_\Sigma}\over{u}}\rfloor \omega _\tau =1\,.\label{uomega}
\end{equation}
The simplest 1-form $\omega _\tau$ consistent with (\ref{uomega}), with (\ref{time04}) at sight, is
\begin{equation}
\omega _\tau =-\,{1\over{(\hat{u}^0)^2}}\,\,u_\mu\,\vartheta _{_\Sigma}^\mu\,.\label{omegatau}
\end{equation}
Then one imposes the Frobenius foliation condition
\begin{equation}
\omega _\tau\wedge d\,\omega _\tau =0\,,\label{Frobenius}
\end{equation}
which constitutes the necessary and sufficient condition for the existence of hypersurfaces orthogonal to the vector field $u$. Eq.(\ref{Frobenius}) induces both, a foliation of the tangent space into simultaneity slices providing a synchronization criterion, and moreover, the existence of a certain time function $\tau$ such that the solution of (\ref{Frobenius}) reads
\begin{equation}
\omega _\tau = N\,d_{_\Sigma }\tau\,.\label{ndtau}
\end{equation}
In terms of (\ref{ndtau}), condition (\ref{uomega}) takes the form
\begin{equation}
{\buildrel {_\Sigma}\over{u}}\rfloor \left( N d_{_\Sigma}\tau\right) =1\,,\label{orthogconditmod}
\end{equation}
so that, in view of (\ref{variety08}) with (\ref{omegatau}), implying ${\buildrel {_\Sigma}\over{u}}$ to be determined up to arbitrary contributions proportional to ${\buildrel {_\Sigma}\over{X}}_i$, the general form of the vector field (\ref{time02}) can be written in the quasi standard form
\begin{equation}
{\buildrel {_\Sigma}\over{u}} = {1\over N}\,\left( \pi _{_{P\Sigma}*}\,\partial _\tau + N^i\,{\buildrel {_\Sigma}\over{X}}_i\,\right)\,,\label{partialtau}
\end{equation}
involving both a lapse and a shift-like function. Comparing (\ref{partialtau}) with (\ref{time03}) and (\ref{variety10}), we find ${1\over N}\,\partial _\tau =u^\mu\,\partial _{\xi ^\mu}$ and $N^i =N\hat{u}^0\,\hat{e}_0{}^i =N u^\mu e_\mu{}^i $. According to (\ref{partialtau}) together with ${\buildrel {_\Sigma}\over{u}} = \hat{u}^{_0} {\buildrel {_\Sigma}\over{\hat{e}}}_{_0}$ in (\ref{time03}), the function $\tau$ can be seen as a (parametric) time function measuring the flow of (proper) time of an observer at rest, since $\hat{e}_0\propto\partial _\tau$ up to irrelevant (horizontal) contributions proportional to ${\buildrel {_\Sigma}\over X}_i$. Function $\tau$ labels simultaneous events placed on successive spacelike hypersurfaces of simultaneity along the time flow. It provides a parametrization of proper time, playing the role of a shared time for locally synchronized neighboring reference frames.

The Frobenius foliation condition (\ref{Frobenius}) is equivalent to the condition of vanishing vorticity since, in terms of $\omega _\tau$ given by (\ref{omegatau}), the vorticity, defined in terms of the Christoffel connection (denoted with curly brackets $\{\}$) with the help of the projectors $h_\alpha{}^\beta := \delta _\alpha ^\beta + u_\alpha u^\beta/(\hat{u}^0)^2$ as in pg. 82 of \cite{Hawking:1973uf} or pg. 70 of \cite{Stephani:2003tm},  reads
\begin{eqnarray}
\omega _{\alpha\beta} &:=& h_{[\alpha}{}^\mu h_{\beta ]}{}^\nu e_\mu\rfloor D^{\{\}}u_\nu \nonumber\\
&=& e_{[\alpha}\rfloor D^{\{\}}u_{\beta ]} +{1\over{c^2}}\,u_{[\alpha}{\cal \L\/}^{\{\}}_u u_{\beta ]}\nonumber\\
&=& {{(\hat{u}^0)^2}\over{2}}\,e_\alpha\rfloor e_\beta\rfloor u\rfloor (\,\omega _\tau\wedge d\,\omega _\tau\,)\,.\label{vorticity}
\end{eqnarray}
(See (\ref{covLiederiv2}) and (\ref{Schwarz06}) and compare with (\ref{Frobenius}).) By defining simultaneity for a congruence of worldlines with the condition to be {\it irrotational}, one aspires to avoid the difficulties in the characterization of simultaneity for relatively rotating frames.

\subsection{Time connection defining simultaneity slices}

Locally, a congruence of curves resembles a fibred space. On the other hand, we have seen that a local definition of simultaneity requires the introduction of local spacelike slices of simultaneity (orthogonal to the fibres of such space) which can be regarded as horizontal hypersurfaces, so that, as pointed out by Minguzzi \cite{Minguzzi:2001ts}--\cite{Minguzzi:2005ag}, they are to be associated with a so called {\it simultaneity connection}. He assigns this role to $\omega _\tau$. We claim that the dependence of local simultaneity on a connection actually constitutes a feature of the bundle structure \cite{Tresguerres:2002uh} supporting the present paper. Indeed, let us show that the one-form (\ref{omegatau}), defining horizontality and thus simultaneity, is the time-translational component of a Poincar\'e connection of the type considered by us.

Evaluating (\ref{omegatau}) referred to a frame where $\hat{u}^a =0$, $\hat{u}^0 \neq 0$, the gauge invariant expression $u_\mu \vartheta _{_\Sigma}^\mu$ reduces to $\hat{u}_0\,\hat{\vartheta}_{_\Sigma}^0$, see (\ref{hattheta0}) with (\ref{timecomp}) and (\ref{spacecomps}), so that 
\begin{equation}
\omega _\tau ={1\over{\hat{u}^0}}\,\hat{\vartheta}_{_\Sigma}^0\,.\label{omegazero}
\end{equation}
According to (\ref{omegazero}), the one-form $\omega _\tau $ coincides (up to multiplicative factors) with the gauge invariant time component (\ref{hattheta0}) --with zero variation (\ref{App2.19a})-- of the translational connection of PGT corresponding to the coframe with $\hat{u}^a =0\,$, so that (\ref{Frobenius}) is equivalent to the (gauge invariant) foliation condition $\hat{\vartheta}_{_{\Sigma}}^0\wedge d\hat{\vartheta}_{_{\Sigma}}^0 =0$ of Ref.\cite{Lopez-Pinto:1997aw}.

The connection 1-form $\hat{\vartheta}_{_{\Sigma}}^0$ defines in particular the horizontality of the spacelike vector fields ${\buildrel {_\Sigma}\over{\hat{e}}}_a$, satisfying the condition ${\buildrel {_\Sigma}\over{\hat{e}}}_a\rfloor\hat{\vartheta}_{_{\Sigma}}^0 =0$. See Fig.4. This role played by $\hat{\vartheta}_{_{\Sigma}}^0$ in the definition of (horizontal) hypersurfaces of simultaneity can be extended to the time component of arbitrary translational connections (\ref{connform4}). So, the time connection $\vartheta _{_{\Sigma}}^0$, boosted with respect to $\hat{\vartheta}_{_{\Sigma}}^0$ (see (\ref{hattheta0})--(\ref{hatthetaa})) defines the horizontality of the spatial basis vectors ${\buildrel {_\Sigma}\over{e}}_a$, and thus the simultaneity of all events contained in the hypersurfaces expanded by them, since also in this case ${\buildrel {_\Sigma}\over{e}}_a\rfloor\vartheta _{_{\Sigma}}^0 =0\,$.
\begin{figure}[h]
\begin{center}
\includegraphics[totalheight=63mm,width=89mm]{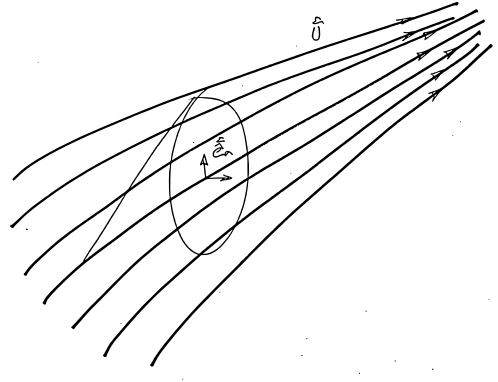}
\caption{\label{Fig4}Simultaneity slices}
\end{center}
\end{figure}

The equality between (\ref{ndtau}) and (\ref{omegazero}) implies the alignment of $\hat{\vartheta}_{_{\Sigma}}^0$ with $d_{_\Sigma}\tau$. Such proportionality $\hat{\vartheta}_{_{\Sigma}}^0\propto d_{_\Sigma}\tau$, in correspondence with the one $\hat{e}_0\propto\partial _\tau$ already mentioned when commenting (\ref{partialtau}), derives from the form of (\ref{omegazero}), resulting from the choice of a frame with zero three-velocity. In both cases, the absence of motion in the appearence of the time function $\tau$ parametrizing the time flow suggests to interpret it as proper time.

\subsection{Bundle aspects of the timelike vector field $u$}

The underlying composite bundle structure where the vector field (\ref{time02}) is defined enables interpretations of its components $u^\mu$ in terms of several preexisting bundle fields such as translational and boost Goldstone fields $\xi ^\mu$ and $\beta ^a$ respectively (together with connections). In the following we discuss the internal structure of $u^\mu$ derived in a natural manner from its relations with those fields.

Taking into account eq.(\ref{5dpoinc80}), ruling the displacement induced on the position object ${\buildrel {_\Sigma}\over p}$ by the operator $\nabla\,$, we introduce the evolution operator $\nabla _u\,$, whose action on ${\buildrel {_\Sigma}\over p}$  we define, in analogy to (\ref{5dpoinc64}), as
\begin{equation}
\nabla _u {\buildrel {_\Sigma}\over p} := \bigl(\,{\buildrel {_\Sigma}\over u}\rfloor\vartheta _{_\Sigma}^\mu\bigr){\buildrel {_\Sigma}\over e}_\mu\,,\label{vel01}
\end{equation}
showing the effect of carrying ${\buildrel {_\Sigma}\over p}$ from fiber to fiber along the flow generated by ${\buildrel {_\Sigma}\over u}$. According to (\ref{5dpoinc81}),(\ref{time02}) and (\ref{time02bis}), the quantity appearing in the r.h.s. of (\ref{vel01}) is simply
\begin{equation}
\bigl(\,{\buildrel {_\Sigma}\over u}\rfloor\vartheta _{_\Sigma}^\mu\bigr) =\bigl(\,{\buildrel {_M}\over u}\rfloor\vartheta _{_M}^\mu\bigr) = u^\mu\,.\label{vel02}
\end{equation}
Then, recalling the tetrad structure (\ref{connform4}) or (\ref{variety06bis}) and definition of ordinary Lie derivatives of forms $\alpha$ with respect to $u$ in its role as {\it time vector} (\ref{partialtau}), namely
\begin{equation}
{\it{l}}_u\alpha :=\,u\rfloor d\alpha + d\,(u\rfloor\alpha\,)\,,\label{Liederdef}
\end{equation}
and their covariant generalization \cite{Hehl:1995ue}, given by
\begin{equation}
{\cal \L\/}_u\alpha ^A:=\,u\rfloor D\alpha ^A + D\,(u\rfloor\alpha ^A\,)\,,\label{covLiederiv2}
\end{equation}
we find
\begin{eqnarray}
u^\mu &=& {\cal \L\/}_u\,\xi ^\mu + {\buildrel (T)\over{\Gamma _{\bot}^\mu}}\nonumber\\
&=& {\it{l}}_u\,\xi ^\mu +\Gamma _{\bot\nu}{}^\mu\,\xi ^\nu + {\buildrel (T)\over{\Gamma _{\bot}^\mu}}\,,\label{fourvelocity}
\end{eqnarray}
along the flow generated by $u$ (see \cite{Tresguerres:2007ih}), where we use the notation $\Gamma _{\bot\nu}{}^\mu := u\rfloor\Gamma _\nu{}^\mu $ and ${\buildrel (T)\over{\Gamma _{\bot}^\mu}} := u\rfloor {\buildrel (T)\over{\Gamma ^\mu}}$. Eq. (\ref{fourvelocity}) equips the vector field components $u^\mu$ with a structure, according to which, changes of ${\buildrel {_\Sigma}\over p}$ as represented by ${\buildrel {_\Sigma}\over u}$ become expressed in terms of bundle fields, in particular in terms of the coordinate-like quantities $\xi ^\mu$ and connections. From (\ref{fourvelocity}) we read out that the measurable relative position vector $\xi ^\mu$ evolves with respect to parametric time, while contributions due to the change of frame (including the origin, see (\ref{5dpoinc78}) and (\ref{5dpoinc79})) ensure the covariance of $u^\mu$.

A second interpretation of $u^\mu$ is possible --compatible with (\ref{fourvelocity}) provided $u$ is timelike-- in terms of Goldstone fields different from the translational parameters $\xi ^\mu$, namely in terms of the boost Goldstone fields (\ref{App2.9}) introduced in \cite{Lopez-Pinto:1997aw} \cite{Tiemblo:2005js} and briefly recalled in Appendix C. Acting with ${\buildrel {_\Sigma}\over{u}}$ on the coframes (\ref{hattheta0}) and (\ref{hatthetaa}) involving those Goldstone fields (\ref{App2.9}), we get
\begin{eqnarray}
\hat{u}^0 &=& \gamma\left(\,u^0 - \beta _a u^a\,\right)\,,\label{hatfourvel0}\\
\hat{u}^a &=& u^a +(\gamma -1){{\beta ^a\beta _b}\over{\beta ^2}}\,u^b -\gamma\beta ^a u^0\,.\label{hatfourvela}
\end{eqnarray}
Then, being the vector field $u$ timelike, we can take as before $\hat{u}^a =0$, $\hat{u}^0\neq 0$, so that (\ref{hatfourvel0}), (\ref{hatfourvela}) imply
\begin{eqnarray}
u^0 &=& \hat{u}^{_0}\gamma \,,\label{timecomp}\\
u^a &=& \hat{u}^{_0}\gamma\beta ^a\,,\label{spacecomps}
\end{eqnarray}
establishing a relationship between the vector field components $u^\mu $ and the Goldstone fields $\beta ^a$ (the latter ones being isomorphic to the boost parameters of the Lorentz group). From (\ref{timecomp}) and (\ref{spacecomps}) with the Minkowski metric (\ref{5dpoinc03}) follows the condition (\ref{time04}) corresponding to a timelike vector whose worldlines occur inside the light cone. The quantity $\hat{u}^0 := {\buildrel {_\Sigma}\over u}\rfloor\hat{\vartheta}_{_\Sigma}^0\,$ is invariant (although not necessarily constant) since it is defined as the contraction of ${\buildrel {_\Sigma}\over u}$ with the invariant time connection (\ref{hattheta0}) with zero variation (\ref{App2.19a}). If desired, one can identify $\hat{u}^0$ with the (constant) velocity of light $c$.

\subsection{Geodesic motion equations}

In the present approach, motion results from the displacement of the position object ${\buildrel {_\Sigma}\over{p}}\,$ along the flow of events, as induced by the evolution operator $\nabla _u$. Otherwise we follow the standard view, assuming the motion of test particles to be geodesic, with the timelike vector field ${\buildrel {_\Sigma}\over{u}}$ satisfying $u_\mu u^\mu =-(\hat{u}^0)^2$. (Here we do not pay attention to the geodesic motion of free propagating light rays involving null vector fields.)

Let us consider the tangent space $T_{( x\,,\xi )}(\Sigma )$ at a given point $(x\,,\xi)\in \Sigma\,$. Positions on it are described by the gauge-theoretical variable ${\buildrel {_\Sigma}\over{p}} = {\buildrel {_\Sigma}\over{\mathfrak{o}}} + \xi ^\mu\,{\buildrel {_\Sigma}\over{e}}_\mu\,$, so that, with respect to the frame $({\buildrel {_\Sigma}\over{e}}_\mu\,,l^{-1}\mathfrak{o})\,$, a position is fully determined by measuring the vector components $\xi ^\mu$. Moreover, for an observer at rest on a given frame, motion manifests itself as a succession of different values of $\xi ^\mu$ of a moving body with respect to the (seemingly) fixed frame. That is, for the observer describing motion, only the relative-position components $\xi ^\mu$ and their alterations are immediately measurable.

However, although not evident to the observer, changes also affect the reference frame. When ${\buildrel {_\Sigma}\over{p}}$ is carried by the flow generated by the timelike vector field ${\buildrel {_\Sigma}\over u}$ from $T_{(x\,,\xi)}(\Sigma )$ to a neighboring fiber at the $\Sigma$-point $(x\,,\xi) + d(x\,,\xi)\,$, then, according to (\ref{time02}),(\ref{vel01}),(\ref{vel02}), the position points experience the change
\begin{equation}
\nabla _u {\buildrel {_\Sigma}\over p} = {\buildrel {_\Sigma}\over u} = u^\mu {\buildrel {_\Sigma}\over e}_\mu\,,\label{vel01bis}
\end{equation}
with the fourvelocity $u^\mu$ given by (\ref{fourvelocity}), the latter containing not only the observable modification ${\it{l}}_u\xi ^\mu$ of the position parameters, but also connection contributions due to the changes of the moving frame, that is, those (\ref{5dpoinc78}) of the vector basis and (\ref{5dpoinc79}) of the origin (all of them referred to the {\it seemingly fixed} frame whose basis vectors ${\buildrel {_\Sigma}\over e}_\mu $ are displayed in (\ref{vel01bis})).

Let us now turn attention to the equation of motion for test particles. As already mentioned, we take it to be the geodesic equation, written in terms of the action of $\nabla _u $ on ${\buildrel {_\Sigma}\over p}$ as
\begin{eqnarray}
\nabla _u \nabla _u {\buildrel {_\Sigma}\over p} &=& \nabla _u {\buildrel {_\Sigma}\over u}\nonumber\\
&=& {\it{l}}_u u^\mu\,{\buildrel {_\Sigma}\over e}_\mu + u^\mu \nabla _u {\buildrel {_\Sigma}\over e}_\mu\nonumber\\
&=& {\cal \L\/}_u u^\mu \,{\buildrel {_\Sigma}\over e}_\mu\nonumber\\
&=& 0\,,\label{eqsmot}
\end{eqnarray}
(which can be seen as an additional condition, together with (\ref{time04}) and (\ref{uomega}), characterizing the vector field ${\buildrel {_\Sigma}\over{u}}$). From (\ref{eqsmot}) we get the law
\begin{equation}
{\cal \L\/}_u u^\mu =0\,,\label{noaccel}
\end{equation}
(also relative to a {\it seemingly fixed} frame ${\buildrel {_\Sigma}\over e}_\mu $) where the covariant acceleration reads
\begin{eqnarray}
{\cal \L\/}_u u^\mu &:=& {\it{l}}_u u^\mu +\Gamma  _{\bot\nu}{}^\mu\,u^\nu \nonumber\\
&=& {\cal \L\/}_u\,{\cal \L\/}_u\,\xi ^\mu + {\cal \L\/}_u\,{\buildrel (T)\over{\Gamma _{\bot}^\mu}}\,.\label{acceleration}
\end{eqnarray}
(The second form of (\ref{acceleration}) derives from (\ref{fourvelocity}).) In virtue of (\ref{noaccel}) with definition (\ref{acceleration}), the connections $\Gamma _{\bot\nu}{}^\mu\,$ and ${\buildrel (T)\over{\Gamma _{\bot}^\mu}}$, induced by the evolution of the basis vectors ${\buildrel {_\Sigma}\over{e}}_\mu $ and of the origin $\mathfrak{o}$ respectively, become observable as a result of their influence on the acceleration ${\it{l}}_u {\it{l}}_u\xi ^\mu$ of the position vector -- as {\it forces} deviating motion from being rectilinear--. In the absence of connections, (\ref{noaccel}) reduces to the inertial law ${\it{l}}_u {\it{l}}_u\xi ^\mu =0$.

It should be clear from the previous exposition that observable motion is not that of the point ${\buildrel {_\Sigma}\over{p}}\,$ as such, but that of the vector components $\xi ^\mu$ relatively to the frame $({\buildrel {_\Sigma}\over{e}}_\mu\,,l^{-1}\mathfrak{o})\,$ when ${\buildrel {_\Sigma}\over{p}}\,$ is displaced (that is, when it {\it evolves}) from fiber to fiber in the direction pointed out by the timelike vector field ${\buildrel {_\Sigma}\over u}$. See Fig.5. Indeed, since no global frame (and in particular, no global origin) exists, motion has to be evaluated locally with respect to a frame evolving on the underlying bundle structure as $\nabla _u ({\buildrel {_\Sigma}\over{e}}_\mu\,,l^{-1}\mathfrak{o})$, see (\ref{5dpoinc78}) and (\ref{5dpoinc79}), where connections are previously determined by field equations (that is, either by the Einstein equation of GR or by its generalizations of gauge theories of gravity).
\begin{figure}[h]
\begin{center}
\includegraphics[totalheight=63mm,width=89mm]{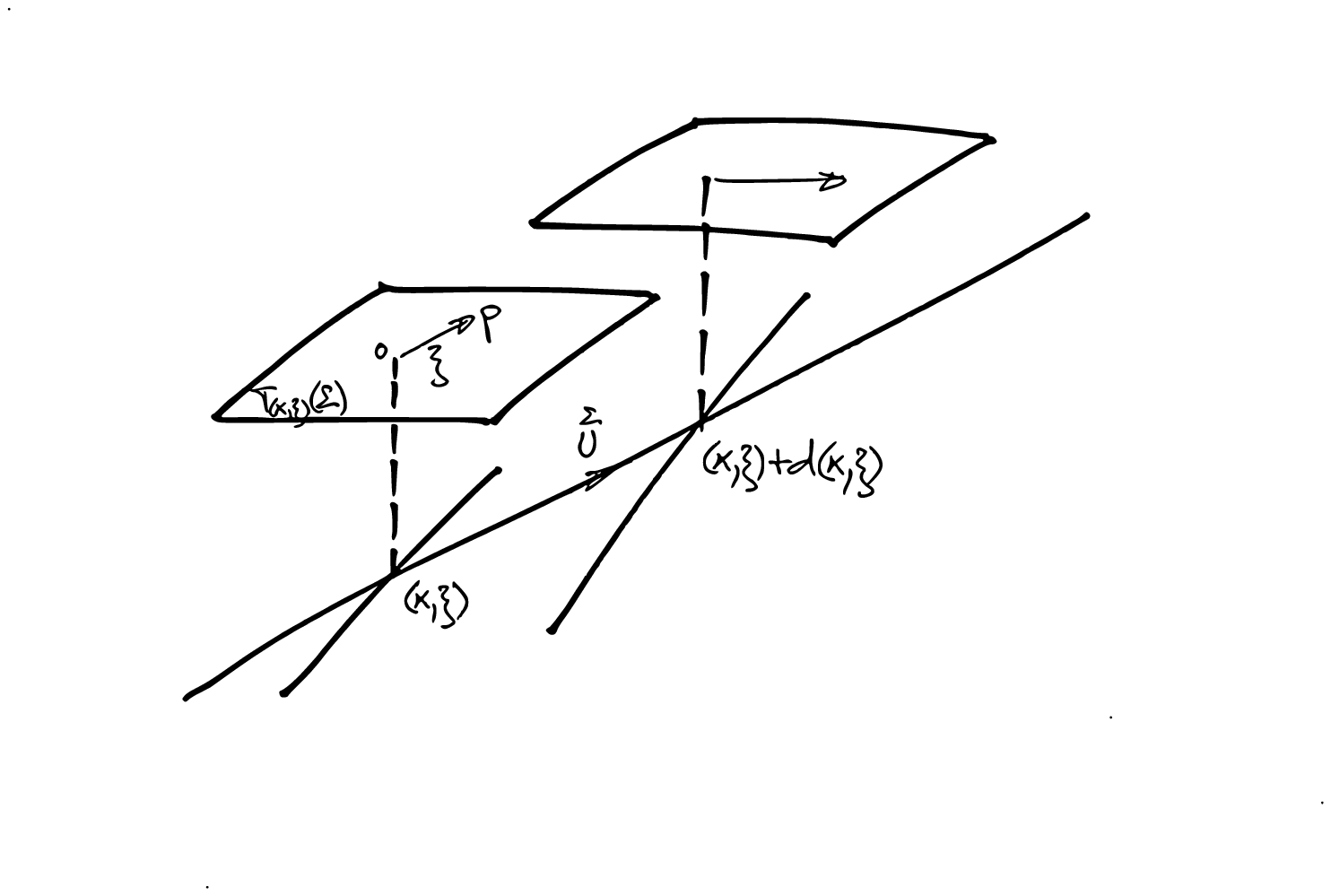}
\caption{\label{Fig5}Underlying time evolution on the bundle}
\end{center}
\end{figure}

Let us finally say that our restriction to timelike vector fields $u$, associated to observers, is justified merely in order to guarantee the interpretation of the function $\tau$ as parametric (proper) time. However, by replacing $\tau$ by a parameter $\lambda$ with no time character, also null vectors associated to light rays are admissible, so that the geodesic motion is determined in general by condition $\nabla _u \nabla _u {\buildrel {_\Sigma}\over{p}} =0$ together with either $u_\mu u^\mu = -(\hat{u}^0)^2$ or $u_\mu u^\mu = 0$, corresponding to massive free point particles or to light rays respectively.

\section{Schwarzschild orbits of test particles}

In the present paper we are not concerned with gravitational dynamics. Nevertheless, we make use of Schwarzschild's spacetime, which is obtained as a solution of Einstein's equations \cite{Weinberg} or of an equivalent gauge theoretical formulation in PGT which can be found elsewhere \cite{Hehl:1995ue} \cite{Obukhov:2006ge} \cite{Tresguerres:2007ih}. We make use of this particular solution in order to illustrate the kind of motion description presented in previous section with the example of a test particle orbiting on a well known geometrical background. In this case, the geodesic evolution of ${\buildrel {_\Sigma}\over{p}} = {\buildrel {_\Sigma}\over{\mathfrak{o}}} + \xi ^\mu\,{\buildrel {_\Sigma}\over{e}}_\mu\,$ according to (\ref{eqsmot}) gives rise to the motion of $\xi ^\mu$ with respect to the (apparently) {\it fixed} frame $({\buildrel {_\Sigma}\over{e}}_\mu\,,l^{-1}\mathfrak{o})\,$, resembling the behavior of a two body system in classical mechanics, with the changing position $\xi ^\mu$ showing the motion of a reduced mass with respect to a center of mass placed at the origin. Fig.6 shows what an observer actually {\it sees} on the tangent space when the latter evolves on the bundle background as depicted in Fig.5. The values of $\xi ^\mu$, referred to the moving frame taken as a {\it fixed} screen, are observed to change according to (\ref{noaccel}) with (\ref{acceleration}).
\begin{figure}[h]
\begin{center}
\includegraphics[totalheight=63mm,width=89mm]{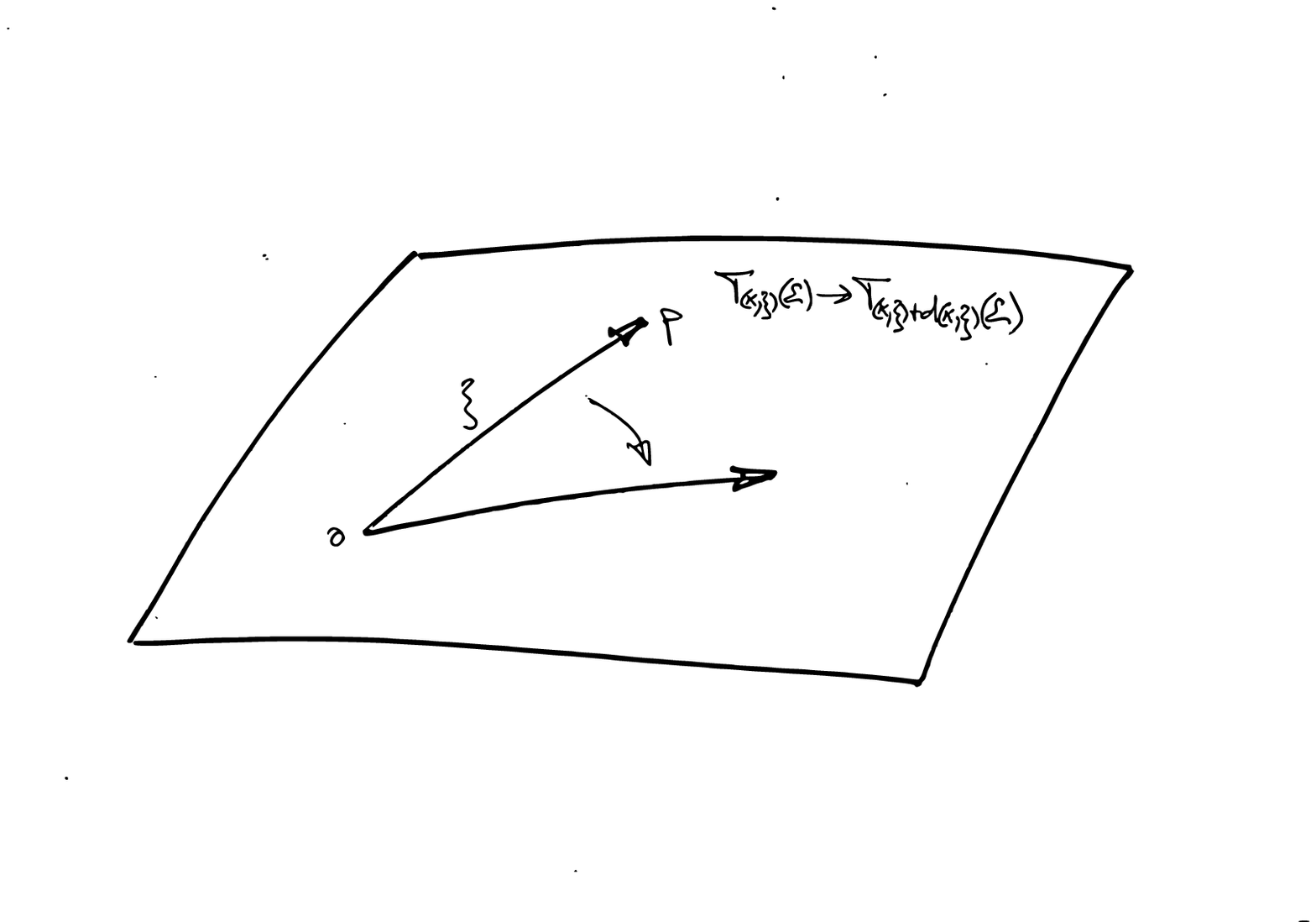}
\caption{\label{Fig6}Observable time evolution}
\end{center}
\end{figure}

\subsection{Schwarzschild tetrads and connections}

The treatment we are going to develop is not univocally determined. However, with our particular choices, it should be illustrative of the general features one can find in the analysis of any motion. Let us consider the Schwarzschild metric solution of General Relativity
\begin{equation}
ds^2 =-\Phi ^2\,c^2 dt^2 +{{dr^2}\over{\Phi ^2}} + r^2\,d\theta ^2 + r^2\sin{\theta}^2\,d\varphi ^2\,,\label{Schwarz01}
\end{equation}
with the Schwarzschild function
\begin{equation}
\Phi :=\sqrt{1-{{2GM}\over{c^2 r}}}\,.\label{Schwarz02}
\end{equation}
In Cartesian coordinates, (\ref{Schwarz01}) takes the form
\begin{equation}
ds^2 =-\Phi ^2\,c^2 dt^2 + \Bigl[\,\delta _{ab}+\Bigl( {1\over{\Phi ^2}}-1\Bigr){{x_a x_b}\over{r^2}}\,\Bigr] dx^a dx^b\,,\label{Schwarz03}
\end{equation}
which can be expressed in terms of tetrads as
\begin{equation}
ds^2 = o_{\alpha\beta}\,\,\vartheta _{_M}^\alpha\,\vartheta _{_M}^\beta\,,\label{Schwarz03bis}
\end{equation}
with the Minkowski metric $o_{\alpha\beta}= diag (-+++)\,$, where we choose
\begin{eqnarray}
\vartheta _{_M}^0 &=& \Phi\,c\,dt\,,\label{Schwarz04}\\
\vartheta _{_M}^a &=& dx^a +\Bigl( {1\over \Phi} -1\Bigr) {{x^a x_b}\over{r^2}}\,dx^b\,,\label{Schwarz05}
\end{eqnarray}
being $r =\sqrt{x_a x^a} =\sqrt{\delta _{ab}\,x^a x^b}\,$. In the absence of torsion, the Lorentz connection on $M$ reduces to the Christoffel connection
\begin{equation}
\Gamma ^{\{\}}_{\alpha\beta}:=\, {\buildrel {_M}\over{e}}_{[\alpha }\rfloor d\,{\buildrel {_M}\over{\vartheta}}_{\beta ]} -{1\over2} \left( {\buildrel {_M}\over{e}}_\alpha\rfloor {\buildrel {_M}\over{e}}_\beta\rfloor d\,\vartheta _{_M}^\gamma\right){\buildrel {_M}\over{\vartheta}}_\gamma \,.\label{Schwarz06}
\end{equation}
(See \cite{Hehl:1995ue}.) In order to calculate (\ref{Schwarz06}) for the Schwarzschild case, from (\ref{Schwarz04}) and (\ref{Schwarz05}) we find
\begin{eqnarray}
d\vartheta _{_M}^0 &=& \partial _r\Phi\,{{x_b}\over r}\,\vartheta _{_M}^b\wedge\vartheta _{_M}^0\,,\label{Schwarz04bis}\\
d\vartheta _{_M}^a &=& \bigl(\Phi -1\bigr) {{x_b}\over{r^2}}\,\vartheta _{_M}^b\wedge\vartheta _{_M}^a\,,\label{Schwarz05bis}
\end{eqnarray}
where
\begin{equation}
\partial _r\Phi :={{GM}\over{\Phi c^2 r^2}}\,,\label{Schwarz07}
\end{equation}
so that the Christoffel components (\ref{Schwarz06}) with (\ref{Schwarz04bis}) and (\ref{Schwarz05bis}) read
\begin{eqnarray}
\Gamma ^{\{\}}_{0a} &=& \Phi\,\partial _r\Phi\,{{x_a}\over{r}}\,c\,dt\,,\label{Schwarz08}\\
\Gamma ^{\{\}}_{ab} &=&  (\Phi -1 ) {{2}\over{r^2}}\,x_{ [ a} dx_{b ]}\,.\label{Schwarz09}
\end{eqnarray}
With (\ref{Schwarz08}) and (\ref{Schwarz09}) at hand, the tetrads (\ref{Schwarz04}), (\ref{Schwarz05}) can be brought to their fundamental form
\begin{equation}
\vartheta _{_M}^\mu = D\xi ^\mu + {\buildrel (T)\over{\Gamma\,^\mu}}\,,\label{Schwarz10}
\end{equation}
established in (\ref{variety06bis}), by identifying the translational parameters $\xi ^\mu$ as
\begin{equation}
\xi ^0 := \Phi\,c\,t\,,\quad \xi ^a :={{x^a}\over \Phi}\,,\label{Schwarz11}
\end{equation}
and the translational connections as
\begin{eqnarray}
{\buildrel (T)\over{\Gamma\,^0}} &=& -c\,d\,(t\,r)\,\partial _r\Phi = -{{GM\,d\,(t\,r)}\over{\Phi\,c\,r^2}}\,,\label{Schwarz12}\\
{\buildrel (T)\over{\Gamma\,^a}} &=& {{x^a}\over{r}}\,\partial _r\Phi\,\Bigr( {{r\,dr}\over{\Phi ^2}} - \Phi ^2\,c^2 t\,dt\,\Bigr)\nonumber\\
&=& {{GM\,x^a}\over{c^2 r^3}}\,\Bigr( {{r\,dr}\over{\Phi ^3}} - \Phi\,c^2 t\,dt\,\Bigr)\,.\label{Schwarz13}
\end{eqnarray}
Notice, comparing (\ref{Schwarz12}) and (\ref{Schwarz13}) with (\ref{5dpoinc79}), that the displacement of the origin by the operator $\nabla $ gives rise to terms proportional to the mass placed on it, constituting the source of the gravitational field.

\subsection{Motion in Schwarzschild spacetime}

Let us now apply the motion equations of Section VIII E to this particular geometry of spacetime. Contracting (\ref{Schwarz04}) and (\ref{Schwarz05}) with the {\it parametric time vector} ${\buildrel {_M}\over u}$, we get the fourvelocity components of the form (\ref{fourvelocity}), whose values reduce to
\begin{eqnarray}
u^0 &=& \Phi\,c\,\dot{t}\,,\label{Schwarz15}\\
u^a &=& \dot{x}^a +\Bigl( {1\over \Phi} -1\Bigr) {{x^a x_b}\over{r^2}}\,\dot{x}^b\,,\label{Schwarz15bis}
\end{eqnarray}
where we use the simplified notation $\dot{x}^i := {\it{l}}_u x ^i = {\buildrel {_M}\over u}\rfloor dx^i$ for the Lie derivative of $x^i$ with respect to the vector field ${\buildrel {_M}\over u}$.

Next we calculate the corresponding motion equations (\ref{noaccel}) making use of (\ref{Schwarz15}) and (\ref{Schwarz15bis}) and of the connection components (\ref{Schwarz08}), (\ref{Schwarz09}). We get
\begin{eqnarray}
0= {\cal \L\/}_u\,u^0 &=& {1\over \Phi}\,{\it{l}}_u ( \Phi ^2\,c\,\dot{t}\,)\,,\label{Schwarz16}\\
0= {\cal \L\/}_u\,u^a &=& \ddot{x}^a + {{x^a}\over r}\Bigl\{\,\bigl( {1\over \Phi} -1\bigr)\ddot{r}\nonumber\\
&&\hskip1.7cm + {{(1-\Phi)}\over{r}}\left(\,\dot{x}_b \dot{x}^b - \dot{r}^2\right)\nonumber\\
&&\hskip1.7cm +{{\partial _r\Phi}\over{\Phi ^2}}\left[\,( \Phi ^2\,c\,\dot{t}\,)^2 - \dot{r}^2\,\right]\,\Bigr\}\nonumber\\
\,.\label{Schwarz17}
\end{eqnarray}
Eq.(\ref{Schwarz17}) can be simplified by eliminating $\ddot{r}$ from it. Indeed, multiplying (\ref{Schwarz17}) by $x_a$, and taking into account that $x_a x^a =r^2$, $x_a \dot{x}^a =r\dot{r}$  and  $x_a \ddot{x}^a =r\ddot{r}-\left(\,\dot{x}_a \dot{x}^a - \dot{r}^2\right)$, we find
\begin{equation}
\ddot{r}= {{\Phi ^2}\over{r}}\left(\,\dot{x}_b \dot{x}^b - \dot{r}^2\right) -{{\partial _r\Phi}\over{\Phi}}\left[\,( \Phi ^2\,c\,\dot{t}\,)^2 - \dot{r}^2\,\right]\,,\label{Schwarz19}
\end{equation}
which, replaced in (\ref{Schwarz17}), yields
\begin{equation}
0 = \ddot{x}^a + {{x^a}\over r}\Bigl\{\,{{(1-\Phi ^2)}\over{r}}\left(\,\dot{x}_b \dot{x}^b - \dot{r}^2\right)+{{\partial _r\Phi}\over{\Phi}}\left[\,( \Phi ^2\,c\,\dot{t}\,)^2 - \dot{r}^2\,\right]\,\Bigr\}\,.\label{Schwarz20}
\end{equation}
Notice that Eqs. (\ref{Schwarz16}), (\ref{Schwarz20}) --obtained by covariantly deriving the components (\ref{Schwarz15}), (\ref{Schwarz15bis}) of a fourvelocity of the form (\ref{fourvelocity}) as studied in the present paper-- exactly reproduce the standard geodesic equations ${{d^2 x^i}\over{d\tau ^2}}+\Gamma ^i_{jk}{{d x^j}\over{d\tau}}{{d x^k}\over{d\tau}}=0$ formulated in terms of a different {\it fourvelocity} ${{d x^i}\over{d\tau}}$ and of the ordinary Christoffel symbols $\Gamma ^i_{jk} ={1\over 2}\,g^{im}\,(\,\partial _j g_{mk} +\partial _k g_{mj} +\partial _m g_{jk}\,)$ built with the metric $g_{00}=-\Phi ^2$, $g_{ab}=\delta _{ab}+\Bigl( {1\over{\Phi ^2}}-1\Bigr){{x_a x_b}\over{r^2}}$ read out directly from (\ref{Schwarz03}).

What follows is standard manipulation of Schwarzschild geodesic equations \cite{Weinberg}. Eq.(\ref{Schwarz16}) implies
\begin{equation}
\Phi ^2\,c\,\dot{t} = K =const\,.\label{Schwarz18}
\end{equation}
On the other hand, in view of the proportionality $\ddot{x}^a\propto x^a$ established by (\ref{Schwarz20}), the angular momentum per unit mass \cite{Weinberg} defined as
\begin{equation}
J_a := \epsilon _{ab}{}^c\,x^b \dot{x}_c\label{Schwarz21}
\end{equation}
is a conserved quantity whose time derivative trivially vanishes. Thus,
\begin{equation}
J^2 = r^2 (\,\dot{x}_b \dot{x}^b -\dot{r}^2 )\label{Schwarz22}
\end{equation}
is a constant of motion. Replacing (\ref{Schwarz18}) and (\ref{Schwarz22}) in (\ref{Schwarz20}), the latter reduces to
\begin{equation}
0 = \ddot{x}^a + {{x^a}\over r}\Bigl\{\,(1-\Phi ^2){{J^2}\over{r^3}} +{{\partial _r\Phi}\over{\Phi}} (\,K^2 - \dot{r}^2\,)\,\Bigr\}\,.\label{Schwarz23}
\end{equation}
Finally, the relation $o_{\alpha\beta} u^\alpha u^\beta =-c^2$ (where we identify $\hat{u}^0 =c\,$) in terms of (\ref{Schwarz15}) and (\ref{Schwarz15bis}) yields
\begin{equation}
-\Phi ^2 c^2\dot{t}\,^2 +(\,\dot{x}_b\dot{x}^b -\dot{r}\,^2\,)+ {{\dot{r}\,^2}\over{\Phi ^2}} =-c^2\,,\label{Schwarz24}
\end{equation}
which, taking into account once more (\ref{Schwarz18}) and (\ref{Schwarz22}), yields
\begin{equation}
\dot{r}^2 = K^2 - \Phi ^2 \left( c^2 + {{J^2}\over{r^2}}\right)\,.\label{Schwarz25}
\end{equation}
From (\ref{Schwarz23}) and (\ref{Schwarz25}), and making use of (\ref{Schwarz02}) and (\ref{Schwarz07}), we get
\begin{equation}
\ddot{x}^a = -\,{{GM x^a}\over{r^3}}\left( 1 + {{3 J^2}\over{c^2 r^2}}\right)\,,\label{Schwarz26}
\end{equation}
which constitutes the general-relativistic (Schwarzschild) modification of Newton's law of gravity. Exact solutions for it can be found in \cite{Scharf:2011ii}. In order to establish contact with this reference, let us simply indicate the following.

Multiplying (\ref{Schwarz26}) by $x_a$ to get the radial equation, we check that it is identical with the derivative of (\ref{Schwarz25}). Thus, we keep the latter as a first integral. Fixing $\theta = \pi /2\,$, we have $J^2 =r^4\dot{\varphi}^2$ and thus $J =r^2\dot{\varphi}$, so that
\begin{equation}
\dot{r} = {{dr}\over{d\varphi}}\,\dot{\varphi} = {{dr}\over{d\varphi}}\,{J\over{r^2}}\,,\label{Schwarz29}
\end{equation}
which replaced in (\ref{Schwarz25}) yields
\begin{equation}
{{dr}\over{d\varphi}} =\sqrt{ {{( K^2 -c^2 )}\over{J^2}}\,r^4 + {{c^2 r_s}\over{J^2}}\,r^3 -r^2 + r_s\,r}\,,\label{Schwarz30}
\end{equation}
where we make use of the Schwarzschild radius
\begin{equation}
r_s := {{2GM}\over{c^2}}\,.\label{Schwarz28}
\end{equation}
Eq. (\ref{Schwarz28}) coincides with Eq.(2.11) of Ref. \cite{Scharf:2011ii}, where it is solved. Notice that solutions for the coordinates $x^i$ should be replaced in (\ref{Schwarz11}) to get the values of the position components $\xi ^\mu$, the latter ones having a more transparent geometrical meaning.

\section{Conclusions}

Motion occurring in the composite fiber bundle structure \cite{Tresguerres:2002uh} proposed by the author as the geometrical framework for Poincar\'e Gauge Theories was formulated in terms of two main quantities. On the one hand, we constructed the gauge invariant position object ${\buildrel {_\Sigma}\over{p}} = {\buildrel {_\Sigma}\over{\mathfrak{o}}} + \xi ^\mu\,{\buildrel {_\Sigma}\over{e}}_\mu\,$, deriving it from a 5-D matrix representation of the Poincar\'e group. On the other hand, we introduced the vector field ${\buildrel {_\Sigma}\over{u}}$ as the generator of the flow of events. Motion is induced by the evolution operator $\nabla _u$ acting on ${\buildrel {_\Sigma}\over{p}}$.

In the course of our study, we recognized different meanings of the vector field ${\buildrel {_\Sigma}\over{u}}$. Its double aspect either as fourvelocity or as proper time vector (that is, time evolution as experienced either by an observer who sees the motion of a body, or by another one moving with the body) is reflected in (\ref{time03}), while (\ref{partialtau}) introduces a (proper) time parametrization for the flow of events. Eq.(\ref{fourvelocity}) expresses the components $u^\mu$ as a Poincar\'e covariant generalization of the Lie derivative of the coordinate-like bundle fields $\xi ^\mu$ with respect to the parametric time vector (\ref{partialtau}), and finally, (\ref{timecomp}) and (\ref{spacecomps}) relate the components $u^\mu$ to the boost-like fields (\ref{App2.9}), making apparent the meaning of the latter ones as fourvelocity components.

The geodesic equation $\nabla _u \nabla _u {\buildrel {_\Sigma}\over{p}} =0$ guiding the displacement of ${\buildrel {_\Sigma}\over{p}}$ along the intermediate space $\Sigma $ of the composite bundle determines the motion of the position components $\xi ^\mu$ of ${\buildrel {_\Sigma}\over{p}}$ with respect to a frame whose {\it motion} (relatively to the bundle) manifest itself (through connections) as an accelerating {\it force} for $\xi ^\mu$.

Lastly, we applied our view to the description of Schwarzshild geodesic motion, showing its consistence with the standard approach.

We claim that the ability of the composite-bundle-structure to make possible a description of motion in terms of as natural quantities as the coordinate-like translational parameters $\xi ^\mu$ may be adduced as an argument in support of such structure as the underlying geometry of PGT.

\begin{acknowledgments}
To my father, {\it in memoriam}.
\end{acknowledgments}

\appendix
\section{NLR's in brief}

Let us consider a Lie group $G$ with a subgroup $H$. Given the $G$ group left action
\begin{equation}
\tilde{g}\,' = g\cdot\tilde{g}\label{nlr01}
\end{equation}
on $\tilde{g}\in G$ factorized as
\begin{equation}
\tilde{g} = b\cdot a\,,\quad a\in H\,,\quad b\in G/H\,,\label{nlr02}
\end{equation}
as in (\ref{condit1}), with $a\in H$ such that
\begin{equation}
a' = h\cdot a\,,\quad h\in H\,,\label{nlr03}
\end{equation}
we get
\begin{equation}
g\cdot b = b'\cdot h\,.\label{nlr04}
\end{equation}
Then, given a linear representation field $\varphi $ of $G$, whose transformation under the corresponding group representation schematically reads
\begin{equation}
\varphi\,' = g\cdot\varphi\,,\label{nlr05}
\end{equation}
one can define a nonlinear representation field
\begin{equation}
\psi := b^{-1}\cdot\varphi\,,\label{nlr06}
\end{equation}
transforming under the action of $G$ as
\begin{equation}
\psi\,' = h\cdot\psi\,,\label{nlr07}
\end{equation}
with the same $h\in H$ introduced in (\ref{nlr03}). Accordingly, given the covariant derivative
\begin{equation}
{\cal D}\varphi := \left(\,d + A\,\right)\varphi\,,\label{nlr08}
\end{equation}
where the linear connection transforms as
\begin{equation}
A\,' = g\,\left(\,d + A\,\right) g^{-1}\,,\label{nlr09}
\end{equation}
so that
\begin{equation}
\left( {\cal D}\varphi \right) ' = g\cdot {\cal D}\varphi\,,\label{nlr10}
\end{equation}
one defines the nonlinear connection
\begin{equation}
\Gamma =  b^{-1}\left(\,d + A\,\right) b\label{nlr11}
\end{equation}
transforming as
\begin{equation}
\Gamma\,' = h\left(\,d + \Gamma\,\right) h^{-1}\,,\label{nlr12}
\end{equation}
and in terms of it the nonlinear covariant derivative
\begin{equation}
D\psi := \left(\,d +\Gamma\,\right)\psi = b^{-1}\cdot {\cal D}\varphi\,,\label{nlr13}
\end{equation}
whose transformation
\begin{equation}
\left( D\psi \right)' = h\cdot D\psi\label{nlr14}
\end{equation}
is in accordance with (\ref{nlr07}).

\section{Several gauge transformations for PGT with $H=$Lorentz}

From (\ref{secttrans2}) and (\ref{sigmaxibis}) with $\sigma _{_{MP}}^{-1}(x)\cdot g\cdot\sigma _{_{MP}}(x)=g$, one easily derives the fundamental equation (\ref{nlr04}) for nonlinear realizations \cite{Coleman:1969sm}--\cite{Tiemblo:2005sx}. Eq.(\ref{nlr04}) is more appropriate than (\ref{secttrans2}) for practical calculations. In the case of $G=$ Poincar\'e and $H=$ Lorentz considered in the present paper, we replace in (\ref{nlr04}) the infinitesimal group elements
\begin{equation}
g = e^{i\,\beta ^{\alpha\beta}\Lambda _{\alpha\beta}}\,e^{i\,\epsilon ^\mu P_\mu}\approx 1+ i\,\beta ^{\alpha\beta}\Lambda _{\alpha\beta} + i\,\epsilon ^\mu P_\mu\label{formula02}
\end{equation}
of the Poincar\'e group and
\begin{equation}
h = e^{i\,\mu ^{\alpha\beta}\Lambda _{\alpha\beta}}\approx 1+ i\,\mu ^{\alpha\beta}\Lambda _{\alpha\beta}\label{formula03}
\end{equation}
of the homogeneous Lorentz group, and we parametrize $b$ and $b'$ respectively as
\begin{equation}
b = e^{-i\,\xi ^\mu P_\mu}\,,\label{formula04}
\end{equation}
with finite translational parameters $\xi ^\mu$, and
\begin{equation}
b' = e^{-i\,{\xi '}^\mu P_\mu} = e^{-i\,{(\,\xi ^\mu +\delta\xi ^\mu )} P_\mu}\,.\label{formula05}
\end{equation}
Then, making use of the Poincar\'e commutation relations (\ref{Lor})--(\ref{trans}), with the help of the Hausdorff-Campbell formulas
\begin{eqnarray}
e^{i\,\xi ^\nu P_\nu}\,\Lambda _{\alpha\beta}\,e^{-i\,\xi ^\nu P_\nu} &=& \Lambda _{\alpha\beta} +\xi _{[\alpha} P_{\beta ]}\,,\label{formula06}\\
e^{i\,\xi ^\nu P_\nu}\,P_\mu\,e^{-i\,\xi ^\nu P_\nu} &=& P_\mu\,,\label{formula07}\\
e^{-i\,{(\,\xi ^\nu +\delta\xi ^\nu )} P_\nu} &\approx& e^{-i\,\xi ^\nu P_\nu} +\delta\,e^{-i\,\xi ^\nu P_\nu}\nonumber\\
&=& e^{-i\,\xi ^\nu P_\nu}\left(\,1+e^{i\,\xi ^\nu P_\nu}\,\delta e^{-i\,\xi ^\nu P_\nu}\,\right)\nonumber\\
&=& e^{-i\,\xi ^\nu P_\nu}\left(\,1 -i\,\delta\xi ^\nu P_\nu\,\right)\,,\label{formula08}
\end{eqnarray}
eq.(\ref{nlr04}) yields on the one hand the value
\begin{equation}
\mu ^{\alpha\beta} = \beta ^{\alpha\beta}\label{5dpoinc41}
\end{equation}
for the $H$-parameter, and on the other hand the variations
\begin{equation}
\delta\xi ^\mu =-\xi ^\nu\beta _\nu{}^\mu -\epsilon ^\mu\label{5dpoinc42}
\end{equation}
of the translational parameters $\xi ^\mu$. Observe how the transformations (\ref{5dpoinc42}) resemble those of Cartesian coordinates.

On the other hand, (\ref{varlcon}) applied to (\ref{pbconn04}) with (\ref{formula02}) yields
\begin{equation}
\delta\Gamma ^{\alpha\beta} =D\beta ^{\alpha\beta}\label{varlorconn}
\end{equation}
and
\begin{equation}
\delta {\buildrel (T)\over{\Gamma ^\mu}}=-{\buildrel (T)\over{\Gamma ^\nu}}\beta _\nu{}^\mu +D\epsilon ^\mu\,,\label{vartransconn}
\end{equation}
while (\ref{varnlcon}) applied to (\ref{nlinconn2}) with (\ref{formula03}) and (\ref{5dpoinc41}) gives rise to (\ref{varlorconn}), as before, and to
\begin{equation}
\delta\vartheta _{_M}^\mu =-\vartheta _{_M}^\nu\,\beta _\nu{}^\mu\,,\label{vartetrad}
\end{equation}
which is the variation of (\ref{variety05}) with (\ref{variety06}) (or of (\ref{variety06bis})).

\section{\bf Poincar\'e composite bundle with $H=SO(3)$}

In \cite{Lopez-Pinto:1997aw} and \cite{Tiemblo:2005js}, a nonlinear approach to PGT was proposed with $G=$Poincar\'e and $H=SO(3)$. A bundle interpretation of such alternative treatment of the Poincar\'e group would require to reelaborate several details of the scheme developed in the present paper. We wont do it here, but in order to clarify certain points discussed in the main text we recall how to derive the $SO(3)$ quantities (\ref{hattheta0}), (\ref{hatthetaa})), related to our translational connection (\ref{connform4}) by a gauge-analogous transformation involving Goldstone boost fields rather than gauge group parameters. By decomposing the Lorentz generators $\Lambda _{\alpha\beta}$ into boosts $K_a$ and space rotations $S_a\,$, defined respectively as
\begin{equation}
K_a :=\,2\,\Lambda _{a0}\,,\quad S_a :=-\epsilon _a{}^{bc} \Lambda _{bc}\,,\quad (a=\,1\,,2\,,3)\,,\label{App2.1}
\end{equation}
we apply the nonlinear transformation law (\ref{nlr04}) to the infinitesimal Poincar\'e group elements
\begin{equation}
g = e^{i\,\epsilon ^\mu P_\mu}\,e^{i\,\beta ^{\alpha\beta}\Lambda _{\alpha\beta}}\approx 1+ i\,\beta ^{\alpha\beta}\Lambda _{\alpha\beta} + i\,\epsilon ^\mu P_\mu \,\,\,\in G\,,\label{formula02bis}
\end{equation}
to the $SO(3)$ group elements
\begin{equation}
h =e^{i\,{\bf{\Theta}} ^a S_a}\approx 1+ i\,{\bf{\Theta}} ^a S_a \quad\in H\,,\label{App2.4}
\end{equation}
and to $b\in G/H$ parametrized as
\begin{equation}
b=e^{-i\,\xi ^\mu P_\mu}e^{i\,\lambda ^a K_a}
\,,\label{App2.5}
\end{equation}
being $\xi ^\mu\,$ and $\lambda ^a $ finite coset fields. So we get on the one hand the variation
\begin{equation}
\delta \xi^\mu =-\xi^\nu \beta _\nu {}^\mu -\epsilon ^\mu\,,\label{App2.7}
\end{equation}
showing that the parameters $\xi^\mu $ associated with the translations behave in fact as coordinates. On the other hand, defining from the boost parameters $\lambda ^a$ in (\ref{App2.5}) the fields
\begin{equation}
\beta ^a :=-\,{{\lambda ^a}\over{|\lambda |}} \tanh |\lambda |\,,\qquad \gamma :=\,{1\over{\sqrt{1-\beta ^2}}}\,,\label{App2.9}
\end{equation}
where $|\lambda |:=\,\sqrt{\lambda _a\lambda ^a}$, the quantity $C^\alpha :=\left(\gamma\,,\gamma\,\beta ^a\,\right)\,$ varies as a Lorentz four--vector, namely
\begin{equation}
\delta C^\alpha = -C^\beta\,\beta _\beta{}^\alpha\,.\label{App2.90}
\end{equation}
Making use of (\ref{App2.1}) we introduce for the analogous of (\ref{nlinconn2}) the notation
\begin{eqnarray}
\Gamma _{_M} &=& -i\,(\,\hat{\vartheta}_{_M}^\alpha P_\alpha +\,\hat{\Gamma}^{\alpha\beta}\Lambda _{\alpha\beta}\,)\nonumber\\
&=& -i\,\hat{\vartheta}_{_M}^0 P_0 -i\,\hat{\vartheta}_{_M}^a P_a +i\,X^a K_a +i\,A^a S_a \,.\label{App2.15}
\end{eqnarray}
Lorentz connections are discussed in Refs. \cite{Lopez-Pinto:1997aw} and \cite{Tiemblo:2005js}. Here we emphasize that the new tetrads with hat in (\ref{App2.15}) relate to the tetrads (\ref{connform4}) derived in the present paper as
\begin{eqnarray}
\hat{\vartheta}_{_M}^0 &=& \gamma\left(\,\vartheta _{_M}^0 - \beta _a\vartheta _{_M}^a\,\right)\,,\label{hattheta0}\\
\hat{\vartheta}_{_M}^a &=& \vartheta _{_M}^a +(\gamma -1){{\beta ^a\beta _b}\over{\beta ^2}}\,\vartheta _{_M}^b -\gamma\beta ^a\vartheta _{_M}^0\,,\label{hatthetaa}
\end{eqnarray}
formally resembling a boost gauge transformation, with the main difference that the Goldstone parameters $\lambda ^a$, and thus (\ref{App2.9}) appearing in (\ref{hattheta0})--(\ref{hatthetaa}), are in fact fields of the theory rather than gauge parameters. Consequently, (\ref{hattheta0})--(\ref{hatthetaa}) are definitions of new variables whose transformation properties depend on (\ref{App2.90}). For the quantities defined in (\ref{App2.15}) we find
\begin{eqnarray}
\delta\hat{\vartheta}_{_M}^0 &=&\,0\,, \label{App2.19a}\\
\delta\hat{\vartheta}_{_M}^a &=&\,\epsilon ^a{}_{bc}\,{\bf{\Theta}}
^b\, \hat{\vartheta}_{_M}^c\,, \label{App2.19b}\\
\delta X^a &=&\,\epsilon ^a{}_{bc}\,{\bf{\Theta}} ^b\,X^c\,, \label{App2.19c}\\
\delta A^a &=&-D\,{\bf{\Theta}} ^a\,,\label{App2.19d}
\end{eqnarray}
where ${\bf{\Theta}}^a$ is the $SO(3)$ group parameter introduced in (\ref{App2.4}), with a certain structure \cite{Lopez-Pinto:1997aw} irrelevant for our present purposes. As read out from  (\ref{App2.19a})--(\ref{App2.19d}), besides the Lorenttz connection components, split into an $SO(3)$ connection $A^a$ plus an $SO(3)$ covector 1--form $X^a$, the translational connections become split into an $SO(3)$ covector --the triad $\hat{\vartheta}^a$-- plus an $SO(3)$ singlet --the time component $\hat{\vartheta}^0\,$-- which results to be gauge invariant.

\section{Useful formulas}

In the main text, we repeatedly make use of equation
\begin{equation}
X\rfloor f^*\omega = f_*\,X \rfloor\omega\label{aux01}
\end{equation}
relating, by means of the interior product, pullbacks and pushforwards acting respectively on differential forms $\omega$ and vectors $X$. Together with
(\ref{aux01}), we use the following property of the pullback of composed functions $f$ and $g$: $\,(g\circ f\,)^* =f^* g^*\,$. In particular, from (\ref{compbundle05}) follows $\pi _{_{PM}}^* =\pi _{_{P\Sigma}}^*\,\pi _{_{\Sigma M}}^*$, from (\ref{sigmaxi}), $s_{_{M\Sigma}}^*\,\sigma _{_{\Sigma P}}^* =\sigma _\xi ^*\,$, and from $\pi _{_{P\Sigma}}\circ\sigma _{_{\Sigma P}}=id_{\Sigma}$ we get $\sigma _{_{\Sigma P}}^*\,\pi _{_{P\Sigma}}^*\,=id_{T^*(\Sigma )}$, etc. Analogously, for the pushforward we have $(g\circ f\,)_* =g_* f_*\,$, so that from (\ref{sectdecomp}) we find $s _{_{MP}*}=s _{_{\Sigma P}*}\,s_{_{M\Sigma}*}\,$, etc.

\end{document}